\documentclass{ieeeaccess}
\usepackage[T1]{fontenc}
\usepackage{cite}
\usepackage{amsmath,amssymb,amsfonts}
\usepackage{algorithmic}
\usepackage{algorithm}
\usepackage{graphicx}
\graphicspath{ {./src/} }
\usepackage{xurl}
\usepackage[protrusion=true,expansion=true]{microtype}
\usepackage{textcomp}
\usepackage{bm}
\usepackage{tabularx}
\usepackage{booktabs}
\usepackage{longtable}
\usepackage{needspace}
\usepackage{seqsplit}
\usepackage[hidelinks,breaklinks]{hyperref}
\usepackage{lmodern}
\usepackage[htt]{hyphenat}

\newcommand{\code}[1]{\texttt{\nolinkurl{#1}}}

\makeatletter
\AtBeginDocument{\DeclareMathVersion{bold}
\SetSymbolFont{operators}{bold}{T1}{lmr}{b}{n}
\SetSymbolFont{NewLetters}{bold}{T1}{lmr}{b}{it}
\SetMathAlphabet{\mathrm}{bold}{T1}{lmr}{b}{n}
\SetMathAlphabet{\mathit}{bold}{T1}{lmr}{b}{it}
\SetMathAlphabet{\mathbf}{bold}{T1}{lmr}{b}{n}
\SetMathAlphabet{\mathtt}{bold}{OT1}{pcr}{b}{n}
\SetSymbolFont{symbols}{bold}{OMS}{cmsy}{b}{n}
\renewcommand\boldmath{\@nomath\boldmath\mathversion{bold}}}
\makeatother

\def\BibTeX{{\rm B\kern-.05em{\sc i\kern-.025em b}\kern-.08em
    T\kern-.1667em\lower.7ex\hbox{E}\kern-.125emX}}

\definecolor{accessblue}{cmyk}{0,0,0,1}

\renewcommand{\headerlogo}{}
\renewcommand{\headerlogoall}{}

\renewcommand{\IEEEmembership}[1]{}

\def\thevol{}
\def\theyear{}
\def\footervolfont{\fontsize{0}{0}\selectfont}

\makeatletter
\def\ps@cleanaccess{%
  \gdef\@oddhead{\parbox[t]{\textwidth}{\mbox{}\\[-6.3mm]\mbox{}\nheaderfont\color{accessblue}\headName\hfill\headerlogo{\raisebox{-2pt}{\,}}\color{black}\vspace*{-1.5mm}\par\hrulefill}}%
  \gdef\@evenhead{\parbox[t]{\textwidth}{\mbox{}\\[-6.3mm]\mbox{}\headerlogo\hfill\nheaderfont\color{accessblue}\headName{\raisebox{-2pt}{\,}}\color{black}\vspace*{-1.5mm}\par\hrulefill}}%
  \gdef\@oddfoot{\raisebox{3pt}{%
   \hbox to 0pc{\hbox to \textwidth{\mbox{}\hfill{\footerpagefont\thepage}}}%
  }}%
  \gdef\@evenfoot{\raisebox{3pt}{%
   \hbox to 0pc{\hbox to \textwidth{{\footerpagefont\thepage}\hfill\mbox{}}}%
  }}%
}

\def\ps@titlepage{\ps@cleanaccess}
\def\ps@headings{\ps@cleanaccess}

\AtBeginDocument{
  \let\ps@titlepage\ps@cleanaccess
  \let\ps@headings\ps@cleanaccess
}
\makeatother

\begin{document}
\sloppy

\title{An Adaptive Multi-Layered Honeynet Architecture for Threat Behavior Analysis via Deep Learning}
\author{\href{https://orcid.org/0009-0008-4046-6362}{\includegraphics[scale=0.06]{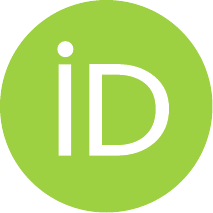}\hspace{1mm}Lukas Johannes Möller}\authorrefmark{1},
\IEEEmembership{Member, IEEE}}
\address[1]{Georgia Institute of Technology Atlanta, United States of America (e-mail: lmoller6@gatech.edu}
\tfootnote{This work was supported in part by the Federal Office for Information Security - BSI}
\markboth
{L J Möller: An Adaptive Multi-Layered Honeynet Architecture for Threat Behavior Analysis via Machine Learning}
{L J Möller: An Adaptive Multi-Layered Honeynet Architecture for Threat Behavior Analysis via Machine Learning}
\begin{abstract}
    The escalating sophistication and variety of cyber threats have rendered static honeypots inadequate, necessitating adaptive, intelligence-driven deception. In this work, ADLAH is introduced: an Adaptive Deep Learning Anomaly Detection Honeynet designed to maximize high-fidelity threat intelligence while minimizing cost through autonomous orchestration of infrastructure. The principal contribution is offered as an end-to-end architectural blueprint and vision for an AI-driven deception platform. Feasibility is evidenced by a functional prototype of the central decision mechanism, in which a reinforcement learning (RL) agent determines, in real time, when sessions should be escalated from low-interaction sensor nodes to dynamically provisioned, high-interaction honeypots. Because sufficient live data were unavailable, field-scale validation is not claimed; instead, design trade-offs and limitations are detailed, and a rigorous roadmap toward empirical evaluation at scale is provided. Beyond selective escalation and anomaly detection, the architecture pursues automated extraction, clustering, and versioning of bot attack chains, a core capability motivated by the empirical observation that exposed services are dominated by automated traffic. Together, these elements delineate a practical path toward cost-efficient capture of high-value adversary behavior, systematic bot versioning, and the production of actionable threat intelligence.
\end{abstract}


\titlepgskip=-21pt

\maketitle
\section{Introduction}
\label{sec:introduction}
\fussy

\subsection{The Evolving Cyber Threat Landscape and Strategic Motivation}
\noindent
The cybersecurity landscape has undergone a dramatic transformation, with cyber threats becoming increasingly sophisticated, persistent, and economically damaging~\cite{enisa2024enisa-658}. Global cybercrime damages are projected to reach \$10.5 trillion annually by 2025~\cite{morgan2020cybercrime-7a9}, driven by the proliferation of connected devices and the professionalization of cybercriminal organizations. Modern attacks are complex, multi-stage campaigns that can persist undetected for extended periods, with a median dwell time of 11 days~\cite{mandiant2025m-trends-03a}.

This research is situated within the strategic context of national cybersecurity agencies, such as the German Federal Office for Information Security (BSI), which are responsible for protecting critical digital infrastructure. A primary goal for such agencies is the rapid detection of novel threats to provide timely warnings. While existing sensor networks like the BSI's MADCAT offer a valuable baseline, the increasing sophistication of application-layer attacks necessitates more advanced analysis capabilities. The Artificial Intelligence (AI)-driven, adaptive architecture presented in this paper aims to address this need, providing a significant leap in capability beyond static sensor grids and directly supporting the BSI's mission to enhance the nation's cyber resilience.~\

\subsection{BSI Collaboration and MADCAT Project Overview}
\noindent
The work presented here has been conducted in collaboration with the German Federal Office for Information Security (BSI) and is designed to complement and extend the BSI's operational capabilities. At the core of the current sensor layer stands MADCAT (Mass Attack Detection Connection Acceptance Tools)~\cite{bsi2025madcat-3e8}, a low-interaction, honeypot-like suite purpose-built to capture Internet-scale, early-stage attack activity for situational awareness and mass-attack analytics.

In recent years, mass attacks on Internet users have increased markedly, with new procedures and attack patterns emerging continuously. Large-scale outages (e.g., in 2016, an incident affected roughly 900,000 consumer connections in Germany) illustrate how mass events can ripple into critical services~\cite{zeit2016hinweise-c6f,spiegel2016bsi-567,schmidt2016grostrung-b5a}. Mass attacks often serve as a staging ground for later targeted operations against specific victims. Many attempts remain undetected or unreported to BSI because they cause no immediately visible impact.

To fulfill its legal mandate under § 7b Abs. 4 BSIG, BSI must therefore operate its infrastructure for collecting and analyzing data on attacks and attempted attacks. Honeypots are a widely used tool for this purpose. By operating its sensors, BSI reduces dependencies on third parties, enables faster and autonomous incident response in significant events, and accumulates longitudinal data to forecast evolving attack trends.

A prominent example is the Log4Shell vulnerability (CVE-2021-44228), disclosed on 10 December 2021.~\cite{zhaojun2021cve-2021-44228-b69} Exploitation attempts are often detectable in the very first network request via malicious JNDI lookups, making them suitable for"First-Flight" analysis. Although the vulnerability is now well known, exploitation remains observable in current traffic, for instance, scanning activity on TCP port 25565 (popular in Minecraft servers) increased noticeably after disclosure and persists to this day.

MADCAT was developed by BSI (C 26) as a universal, low-interaction cyber threat detection suite. It is honeypot-like in that it records all contact attempts without binding itself to specific services."Low interaction" indicates that, in its current v2 deployment, responses are limited to the technically necessary minimum to establish a connection and capture attack vectors. Internally, an evolved version of MADCAT v2 is in operational use. The extensions proposed in this work enable the analysis of all layers above, including session, presentation, and application-layer behavior, thereby supporting scalable prioritization, advanced Layer 7 analytics, and triage across large sensor fleets and dynamically orchestrated honeypots.

Sensors are deployed both at consumer-grade, Internet-reachable access links and on leased servers (e.g., at mass hosters). In controlled test phases, small-scale deployments have recorded on the order of hundreds of thousands of events per day, exhibiting complex and constantly changing mass-attack activity. These observations empirically motivate an adaptive escalation layer that concentrates scarce high-interaction resources on the most valuable sessions.~\footnote{Valuable sessions are defined as interactions that exhibit novel, complex, or persistent behavior, indicating a higher likelihood of yielding actionable intelligence about new threats or adversary TTPs.}

The program serves BSI's statutory tasks (e.g., \S\ 3 Abs. 1 S. 2 Nr. 1 bis 4, 10, 14 BSIG) by building and maintaining expertise in the detection and analysis of attack patterns, and by enabling peer-level information exchange with ISPs, operators, and partner organizations.

In the course of this work, an exchange between Telekom Security, BSI, and the author was conducted to explore whether T-Pot datasets could be made available for research. The contact at Telekom Security indicated that many similar requests are received and that corresponding approval processes can be lengthy. Access could therefore not be granted within the thesis timeline. To mitigate this, a self-hosted T-Pot instance was deployed and used to collect exploratory data. While this analysis is not central to the present results, the resulting dataset is expected to play a role in future iterations of the architecture, particularly for improving application-layer anomaly detection.

MADCAT focuses on recording initial contact attempts across all reachable ports and protocols with minimal response surface, enabling broad deployment at scale (e.g., consumer links and leased servers). In controlled deployments, MADCAT sensors have observed on the order of hundreds of thousands of events per day across a pretty limited number of sensors, underscoring the need for efficient downstream processing.

From an architectural standpoint, MADCAT provides a modular, ingest-ready telemetry layer:
\begin{itemize}
    \item \textbf{Capture Modules (TCP/UDP/ICMP/RAW):} Protocol-specific collectors record first packets and flows. The TCP path includes a postprocessor that correlates SYN and connection objects and optionally leverages system conntrack; its status is summarized via \texttt{ct\_status} (Found/Changed/Failed/None). The RAW path uses libpcap filters to quantify traffic classes (e.g., IPv6)\footnote{Attacks using IPv6 are yet relatively rare compared to IPv4, thus specialized IPv6-capable modules are still under development.} beyond the core protocol set.
    \item \textbf{Enrichment Processor:} Normalizes and augments events (e.g., external sensor IP resolution, geo-IP), performs optional payload splitting for large events using a consistent \texttt{SPLIT} tag, and writes structured logs for shipper ingestion.
    \item \textbf{Monitoring Module:} Provides operational and security telemetry (CPU/memory/disk, process watchdogs, audited executions, listener whitelists) to support secure, reliable field operation of sensors.
\end{itemize}

This paper positions MADCAT as the scalable, low-interaction sensing layer in a larger, adaptive system. The proposed ADLAH architecture learns, in real time, when to escalate a session from a MADCAT Sensor Node to a dynamically provisioned, high-interaction honeypot. In this way, ADLAH preserves MADCAT's breadth and safety while adding depth where it matters.

By driving selective, context-aware escalation decisions, ADLAH provides the following benefits for BSI and MADCAT operations:
\begin{itemize}
    \item increases the intelligence yield per MADCAT event by capturing post-compromise behavior only when warranted,
    \item reduces analyst and compute load in ELK or MongoDB by prioritizing sessions with higher expected value,
    \item supports timely national warnings through earlier identification of novel tools and techniques, and
    \item maintains a small exposure surface on widely deployed Sensor Nodes while concentrating risk within isolated, ephemeral high-interaction backends.
\end{itemize}

This collaboration goal is to deliver an adaptive, ML-driven orchestration layer that directly leverages MADCAT data and infrastructure, thereby enhancing BSI's operational resilience and the strategic value of its sensor network.

\subsection{Limitations of Traditional Honeypot Systems}
\noindent
Traditional honeypot and honeynet systems, while valuable for threat intelligence collection, suffer from several fundamental limitations that reduce their effectiveness against modern threats~\cite{spitzner2002honeypots-ba5}. Static configuration represents perhaps the most significant limitation, as traditional honeypots are typically deployed with fixed configurations that cannot adapt to changing attack patterns or emerging threats.~\footnote{This static nature also extends to the network level.} This static nature makes them increasingly ineffective against sophisticated adversaries who can quickly identify and avoid honeypots through reconnaissance and behavioral analysis, a trend highlighted in the ENISA Threat Landscape report~\cite{enisa2024enisa-658}.

Limited correlation capabilities prevent traditional systems from connecting related attack activities across multiple sessions, periods, or infrastructure components. This limitation is particularly problematic when dealing with advanced persistent threats that employ multi-stage attack campaigns spanning weeks or months. Without the ability to correlate activities across sessions, security analysts miss critical patterns that could reveal the full scope and intent of sophisticated attacks~\cite{project2004know-620}.

Resource inefficiency is another critical limitation, as traditional honeypots often consume significant computational and network resources regardless of the actual threat level or value of collected intelligence. This inefficient resource utilization limits the scale at which honeypots can be deployed and reduces their cost-effectiveness for many organizations. Additionally, traditional systems cannot dynamically adjust their engagement strategies based on adversary behavior, missing opportunities to collect more valuable intelligence from sophisticated threats while wasting resources on low-value automated scanning~\cite{dowling2020new-1b1}.

Detection evasion has become increasingly problematic as adversaries develop more sophisticated techniques for identifying and avoiding honeypots. Modern attack tools often include honeypot detection capabilities that can identify common honeypot signatures, network configurations, and behavioral patterns~\cite{wang2010honeypot-fd5}. Once identified, adversaries avoid these systems, rendering them ineffective for intelligence collection.

\subsection{The Promise of Adaptive Honeynets}
\noindent
Adaptive honeynets represent a paradigm shift in cyber deception technology, offering the potential to overcome the limitations of traditional static systems through intelligent automation and machine learning~\cite{wagener2011adaptive-ce5}. By incorporating Reinforcement Learning (RL) algorithms, adaptive honeynets can learn from their behavior and continuously optimize their deployment of adversary strategies to maximize intelligence collection while minimizing resource consumption~\cite{dowling2018improving-a2d, pauna2018qrassh-db1, veluchamy2022deep-98e}.

The integration of real-time anomaly detection enables adaptive systems to identify novel attack patterns and zero-day exploits that signature-based systems would miss. This capability is particularly valuable in the current threat landscape, where adversaries increasingly rely on custom tools and techniques designed to evade traditional detection methods. Machine learning-based anomaly detection can identify subtle patterns in network traffic, system behavior, and attack sequences that human analysts might overlook~\cite{shone2018deep-dfe, catillo2022autolog-789, park2018anomaly-3c7, vartouni2018anomaly-423}.

Dynamic resource allocation enables adaptive honeynets to automatically scale their deployment according to current threat levels and available resources. This capability ensures that computational resources are focused on the most valuable threats while maintaining cost-effectiveness. During periods of high attack activity, the system can automatically deploy additional honeypot instances to capture more intelligence, while scaling back during quiet periods to conserve resources~\cite{dowling2020new-1b1, boureanu2023honeyiot-893, limouchi2021reinforcement-f6a, pauna2019rewards-e5e}.

Cross-session correlation capabilities enable adaptive honeynets to track sophisticated attack campaigns across multiple sessions, periods, and infrastructure components. This correlation is crucial for comprehending the full extent of advanced persistent threats and devising effective countermeasures. By maintaining persistent adversary profiles and tracking the evolution of adversary behavior over time, adaptive systems can provide valuable insights into adversary motivations, capabilities, and likely future targets~\cite{verkerken2020unsupervised-5db, jadav2021machine-e2c, nguyen2025fednids-6dd}.

Crucially, this research shifts the paradigm of adaptation from the "in-service" level, where a single honeypot alters its responses, to the infrastructure level, where the system orchestrates the deployment of deception resources themselves. This fundamental distinction positions ADLAH as bridging the gap between single-service adaptation and full infrastructure automation.

This shift towards infrastructure-level orchestration provides significant economic and ecological benefits. By selectively deploying resource-intensive, high-interaction honeypots only when a threat is deemed credible, the system radically reduces operational costs. This "smart deployment" approach minimizes baseline computing power and energy consumption, leading to a more sustainable and cost-effective security posture. Furthermore, it lowers the data overhead associated with constant, large-scale logging, focusing analytical resources where they are most needed. This efficiency provides a practical entry point for creating more diverse and intelligent honeypot deployments that can extend beyond SSH to a broader range of services, enhancing the overall strategic capability of the deception infrastructure.
\subsection{Research Objectives and Contributions}
\noindent
This research addresses the critical gap between traditional static honeypot systems and the dynamic, intelligent defense capabilities required to counter modern cyber threats. The primary objective is to develop and demonstrate a practical adaptive honeynet architecture that combines reinforcement learning and dynamic container orchestration, bridging the gap between single-service adaptation and full infrastructure automation.

The research makes several key contributions to the field of cybersecurity:
\begin{enumerate}
    \item It presents a practical prototype of an RL-driven deployment trigger. The system uses a Deep Q-Network (DQN) agent combined with a Long Short-Term Memory (LSTM)~\cite{hwang2019lstm-based-797} to analyze sequences of network events and automatically decides when to escalate a session from a low-interaction sensor to a high-interaction honeypot.
    \item It provides a comprehensive architectural framework for adaptive honeynets that addresses not only technical implementation but also operational, ethical, and legal considerations. This framework serves as a blueprint for organizations seeking to implement adaptive deception technologies.
    \item It contributes to the broader understanding of how machine learning can be effectively applied to cybersecurity challenges, providing insights that extend beyond honeypot systems.
\end{enumerate}

\subsection{Scope and Limitations}
\noindent
This research focuses on the development and demonstration of an adaptive honeynet for network-based threat detection. The scope includes the design of RL algorithms for deployment decision-making, the integration of anomaly detection, and the development of container orchestration for dynamic honeypot deployment. However, it is crucial to acknowledge several significant limitations that define the boundaries of the current work and provide a clear path for future research.

\begin{itemize}
    \item \textbf{Focus on Network-Level Triggers:} The prototype's decision-making is based entirely on first-packet network data. While effective for identifying scans and initial connection attempts, it lacks visibility into encrypted traffic and sophisticated application-layer attacks that may only manifest after a session is established. The system can decide \emph{to deploy} a honeypot, but it cannot analyze the complex, post-compromise behavior within that honeypot to the full extent envisioned in the architecture. ~\footnote{Post-compromise behavior includes lateral movement, privilege escalation, data staging, and tool installation, which are critical for understanding an attacker's ultimate objectives.}

    \item \textbf{Reward Function Sophistication:} The RL agent's effectiveness is fundamentally tied to the quality of its reward function. The prototype employs a temporary and simplistic, quantity-based metric where the reward is calculated based on the amount of logs generated ($N_{\text{logs}}$). This approach creates a potential vulnerability: it can be gamed by a low-sophistication adversary generating high volumes of noise, which the agent might incorrectly value more than a stealthy, advanced adversary making few but critical actions. Future implementations should incorporate more sophisticated metrics, such as anomaly scores, which can better quantify the actual "worth" of an agent's decision by evaluating the novelty and potential significance of the captured interaction. Designing a reward function that accurately quantifies this "intelligence value" by balancing novelty, severity, and relevance remains a major open research challenge~\cite{kristyanto2022evaluation-dbc, innab2018hybrid-396}.

    \item \textbf{Dependency on Live Traffic for Training:} Offline training is only partially applicable. While the agent learns from the consequences of its actions (deploying a honeypot) in a live feedback loop, which complicates pure offline evaluation, methods from Offline RL/Batch RL could be explored with logged interactions and modeled or retrospective rewards. This remains future work; our current prototype assumes live interaction for meaningful policy learning.

    \item \textbf{Vulnerability to Honeypot Detection:} The system's efficacy rests on the assumption that adversaries will interact with the deployed honeypots. However, sophisticated adversaries actively employ anti-honeypot and anti-Virtual Machine (VM)/container detection techniques. They may probe for environmental artifacts, network latency inconsistencies introduced by the redirection mechanism, or predictable behaviors of emulated services to unmask the deception and evade capture~\cite{touch2022comparison-af3}.

    \item \textbf{Scalability and Performance Bottlenecks:} The prototype, while functional, is not designed for large-scale, high-traffic environments. The imperative `iptables` redirection mechanism could become a bottleneck on the sensor node. Furthermore, the computational requirements for the ELK (Elasticsearch, Logstash, Kibana) stack and a fleet of ML models can be substantial, potentially limiting the number of concurrent sessions that can be analyzed and managed, posing a challenge for resource-constrained organizations.

    \item \textbf{Prototype-to-Production Gap:} The current implementation is a prototype designed to demonstrate feasibility, not a production-hardened system. A significant engineering effort would be required to achieve operational readiness, including implementing high-availability and failover mechanisms, comprehensive security hardening of all management components, and developing a more robust, declarative traffic management system to replace the prototype's script-based approach.

    \item \textbf{Ethical and Legal Considerations:} The deployment of honeypots, particularly high-interaction ones, operates in a complex ethical and legal landscape. Issues surrounding entrapment, the collection and storage of potentially sensitive adversary data (including personally identifiable information), and compliance with data protection regulations like the General Data Protection Regulation (GDPR) must be carefully evaluated by any organization deploying such a system within its specific legal jurisdiction~\cite{spitzner2002honeypots-ba5}.
\end{itemize}

\subsection{Document Organization}
\noindent
The remainder of this paper is organized as follows.  Section~\ref{sec:terminology} establishes a clear lexicon by defining the core concepts, technical terms, and evaluation metrics used throughout the paper. Section~\ref{sec:related_work} provides a comprehensive review of foundational research in honeypot technology, machine learning for network defense, and adaptive cyber deception, identifying the key research gap this work addresses. Section~\ref{sec:threat_landscape} offers a detailed analysis of modern cyber threats, including Advanced Persistent Threats (APTs) and ransomware, providing the strategic context for the proposed architecture. Section~\ref{sec:architecture} presents the complete, high-level architecture of the proposed adaptive honeynet, detailing its modular components, data flows, and an AI-driven pipeline for intelligent orchestration. Section~\ref{sec:evasion} discusses the challenge of honeynet detection, reviews state-of-the-art evasion and counter-evasion techniques, and outlines the architecture's approach to achieving stealth. Section~\ref{sec:methodology} details the research methodology, including the experimental environment, the process of data collection from live traffic, and the feature engineering pipeline used to prepare data for the machine learning models. Section~\ref{sec:prototype} describes the practical implementation of the functional prototype, focusing on the reinforcement learning agent, dynamic container deployment, and the traffic redirection mechanism. Section~\ref{sec:evaluation} proposes how to evaluate the system's performance and effectiveness.  The paper concludes in Section~\ref{sec:conclusion} with a summary of key contributions, followed by Section~\ref{sec:future_work}, which outlines a detailed roadmap for future research directions.

\section{Terminology and Definitions}
\label{sec:terminology}

\noindent
This section provides clear definitions of key terms and concepts used throughout this paper to ensure consistent understanding and avoid ambiguity.

\subsection{Core Concepts}

A honeypot is a security mechanism designed to detect, deflect, or study attempts at unauthorized use of information systems~\cite{spitzner2002honeypots-ba5}. Honeypots are decoy systems that appear to be legitimate targets but are isolated and monitored environments designed to attract and analyze malicious activity.

A honeynet is a network of honeypots designed to capture and analyze malicious traffic on a larger scale~\cite{project2004know-620}. Unlike individual honeypots, honeynets offer comprehensive monitoring capabilities across multiple systems, enabling the detection of coordinated attacks that span various targets.

An adaptive honeynet is a honeynet system that can dynamically adjust its configuration, deployment strategy, and response mechanisms based on observed threat patterns and system performance~\cite{wagener2011adaptive-ce5}. Adaptive honeynets utilize machine learning and automation to optimize resource allocation and enhance the effectiveness of threat detection.

Reinforcement learning (RL) is a machine learning paradigm where an agent learns to make decisions by interacting with an environment and receiving feedback in the form of rewards or penalties~\cite{sutton2014reinforcement-c13}. In the context of honeynets, RL agents learn optimal deployment strategies through trial and error.

\subsection{Technical Terms}

A Deep Q-Network (DQN) is a reinforcement learning algorithm that combines Q-learning with deep neural networks to handle high-dimensional state spaces~\cite{mnih2013playing-0cf}. DQN enables the agent to learn complex decision-making policies in environments with large state spaces.

An autoencoder is a neural network architecture designed to learn efficient representations of input data by training the network to reconstruct the input from a compressed latent representation~\cite{vincent2010stacked-165}. In anomaly detection, autoencoders identify unusual patterns by measuring reconstruction error.

Container orchestration refers to the automated management of containerized applications, including deployment, scaling, and monitoring~\cite{merkel2014docker-9f8}. Kubernetes is the primary container orchestration technology referenced in this work; the prototype specifically implements the lightweight k3s distribution, while production use would typically favor managed offerings such as GKE.

A pod is the smallest deployable unit in Kubernetes, representing a group of one or more containers (a "pod") sharing storage and network resources~\cite{bernstein2014containers-35c}. In this context, pods represent individual honeypot instances.

\subsection{Network and Security Concepts}

First-packet data are the initial packets or connection attempts that occur when a session is established~\cite{hwang2020unsupervised-5c4}. First-packet data provide early indicators of adversary intent and behavior before full engagement occurs.

Destination NAT (DNAT) is a network address translation technique that redirects incoming traffic to different destinations~\cite{california1981internet-069}. In honeynet deployments, DNAT is used to redirect adversary traffic to appropriate honeypot instances transparently.

An attack chain is a sequence of related attack actions performed by the same threat actor or group across multiple sessions or timeframes. Attack chains represent coordinated attack campaigns that span various stages and targets~\cite{hutchins2011intelligence-driven-c04}.

Threat intelligence refers to threat information that has been aggregated, transformed, analyzed, interpreted, or enriched to provide the necessary context for decision-making processes~\cite{johnsonnoyearguide-761}. In honeynet contexts, threat intelligence is derived from analysis of adversary behavior, tools, and techniques.

\subsection{Evaluation Metrics}

Detection accuracy is the ability of the system to correctly identify malicious activity while minimizing false positives~\cite{fawcett2006introduction-e3e}. It is measured through precision, recall, F1-score, and area under the ROC curve (AUC).

Resource efficiency is the optimal use of computational resources, including CPU, memory, storage, and network bandwidth~\cite{armbrust2010view-fd7}. Resource efficiency is critical for cost-effective honeynet deployment.

Response time is the time required for the system to detect a threat and initiate appropriate countermeasures~\cite{shneiderman2017designing-a5c}. Low response times are essential for effective threat mitigation.

Threat coverage is the percentage of different attack types and vectors that the system can successfully detect and analyze~\cite{scarfonenoyearguide-6ed}. Comprehensive threat coverage provides broad protection against a wide range of attack methodologies.

\subsection{Operational Terms}

A field trial is a real-world test of the system in an operational environment with actual Internet traffic and genuine adversary interactions~\cite{pauna2014rassh-54e}. Field trials provide validation of system effectiveness under realistic conditions.

Cross-session analysis is the ability to correlate and analyze attack patterns across multiple sessions, timeframes, and network segments~\cite{verkerken2020unsupervised-5db}. Cross-session analysis is essential for detecting sophisticated, multi-stage attacks.

Behavioral profiling is the process of analyzing and categorizing adversary behavior patterns to identify threat actors, attack methodologies, and campaign objectives~\cite{song2010comparative-9e7}. Behavioral profiling enables proactive threat detection and response.

Adaptive deployment is the dynamic allocation of honeypot resources based on real-time threat assessment and system performance~\cite{touch2021asguard-ae5}. Adaptive deployment optimizes resource utilization while maintaining effective threat detection capabilities.

\section{Related Work}
\label{sec:related_work}
\noindent
A comparative analysis of related works is presented in Table~\ref{tab:related_work_comparison}.

\begin{table*}[t]
\caption{Comparative Analysis of Related Work in Anomaly Detection and Adaptive Honeypots\label{tab:related_work_comparison}}
\centering
\renewcommand{\arraystretch}{1.3}
\begin{tabularx}{\textwidth}{l X X X X X}
\hline
\textbf{Work} & \textbf{Primary Goal} & \textbf{Core Methodology} & \textbf{Adaptation Level} & \textbf{Key Contribution} & \textbf{Research Gap Addressed by ADLAH} \\
\hline
\textbf{This Work (ADLAH)} & \textbf{Adaptive Honeynet Orchestration} & \textbf{DQN Reinforcement Learning} & \textbf{Infrastructure-Level} & \textbf{Novel use of RL for dynamic, containerized deployment decisions based on first-packet data.} & \textbf{N/A (Represents the proposed solution)} \\
\hline
RASSH~\cite{pauna2014rassh-54e} & Adaptive Honeypot Engagement & Q-Learning / SARSA & In-Service (Session) & Foundational work applying RL to manipulate SSH sessions to maximize interaction time. & Limited to adapting responses within a single, non-containerized service; no infrastructure orchestration. \\
\hline
Asguard~\cite{touch2021asguard-ae5} & Adaptive Honeypot Engagement & Deep Q-Network (DQN) & In-Service (Session) & Modernizes RL-based honeypots with DQNs for more nuanced interaction and command handling. & Focus remains on single-service engagement, not infrastructure orchestration. \\
\hline
GASH~\cite{prasad2025generative-ccb} & Adaptive Honeypot Engagement & DQN + Generative AI (LLM) & In-Service (Session) & Integrates an LLM (GPT-4o) to generate realistic and convincing responses to attacker commands. & Focuses on improving the quality of interaction within a single service, not on the strategic deployment of resources. \\
\hline
UNADA~\cite{bao2015near-f48} & Autonomous Attack Characterization & Unsupervised Sub-space Clustering & Offline Analysis & Autonomous, near real-time characterization of attacks from NetFlow data without prior knowledge. & Relies on classical ML and aggregated flow data, missing early, packet-level indicators for real-time decisions. \\
\hline
RNN-IDS~\cite{yin2017deep-f0b} & Network Intrusion Detection & Recurrent Neural Network (RNN) & N/A (Classification) & Demonstrates deep learning for general intrusion detection on pre-processed datasets (NSL-KDD). & Designed for classification on static datasets, not for real-time orchestration decisions on live, raw traffic. \\
\hline
Raw Packet Transformers~\cite{sharan2023raw-aab} & Malicious Traffic Classification & Transformer (ByT5) & N/A (Classification) & Proves the feasibility of end-to-end learning directly on raw packet bytes for classification. & High computational cost and focus on classification make it unsuitable for real-time orchestration. \\
\hline
\end{tabularx}
\end{table*}
\subsection{Foundations in Honeypot and Deception Technology}
The concept of a honeypot, a decoy system designed to be probed, attacked, and compromised, has been a cornerstone of defensive security for decades. Early work, famously documented by Clifford Stoll in "The Cuckoo's Egg"~\cite{stoll1989cuckoos-764}, demonstrated the immense value of observing adversaries in a controlled environment to understand their methods. This concept was formalized and popularized by pioneers like Lance Spitzner, whose work laid the groundwork for classifying honeypots based on their level of interaction and purpose (research vs. production)~\cite{spitzner2002honeypots-ba5}.

Low-interaction honeypots, such as the BSI's MADCAT Sensor Node~\cite{bsi2025madcat-3e8} used in this research, emulate services at a superficial level. They are designed to capture initial connection attempts, scans, and basic probes with minimal risk and resource consumption. Their primary advantage is scalability; they can be deployed widely to gather broad statistics on network-level activity. However, they cannot engage sophisticated adversaries or capture complex, post-exploitation behavior.

High-interaction honeypots, in contrast, provide a real, fully functional operating system and services for adversaries to interact with. Systems like Cowrie~\cite{teamnoyearcowrie-975} emulate an SSH and Telnet environment, allowing for the capture of detailed command sequences, tool downloads, and lateral movement attempts. The trade-off is a significantly higher resource footprint and an increased security risk, as the honeypot itself could be used as a staging ground for further attacks if not properly isolated. The concept of a honeynet, a network of multiple honeypots, was introduced by The Honeynet Project to create a more convincing and extensive deception environment, enabling the analysis of more complex, coordinated attacks~\cite{project2004know-620}. Modern honeypot distributions, such as T-Pot~\cite{noauthornoyeart-pot-7d2}, bundle multiple honeypot types into a single platform.

Despite their value, these traditional systems are fundamentally static. They are deployed with a fixed configuration and remain unchanged, making them susceptible to detection by experienced adversaries and inefficient in their use of resources. This static nature is the primary motivation for the shift towards adaptive systems.

A seminal example of this early paradigm is the work by Wagener et al. \cite{wagener2011adaptive-ce5}, often termed "in-service adaptation." Their system utilized a game-theoretic model combined with reinforcement learning to enable a single honeypot to make real-time decisions on how to handle an attacker's commands. The agent's actions were to `allow`, `block`, `substitute`, or `insult` the command, thereby manipulating the interactive session to maximize information gain. While groundbreaking, this approach focuses on optimizing the behavior \textit{within} a single, active honeypot. The architecture presented in this paper, ADLAH, operates at a higher, complementary level of abstraction. Instead of adapting responses to commands, ADLAH adapts the infrastructure itself, using RL to make the strategic decision of \textit{whether} to deploy a high-interaction honeypot in the first place, thus focusing on resource orchestration rather than session manipulation.
\subsection{The Advent of Adaptive Cyber Deception}
The limitations of static honeypots led to research into adaptive systems that could dynamically alter their behavior in response to adversary actions. Early work by Wagener et al. proposed the concept of self-configurable honeypots that could adjust their services and configurations based on observed traffic, laying the conceptual groundwork for the field~\cite{wagener2011adaptive-ce5}.

The integration of machine learning advanced this paradigm. Researchers began to explore how ML could automate the decision-making process, moving beyond simple rule-based adaptations. For example, some systems focused on using ML to improve the believability of the honeypot. Dowling et al. proposed a framework for agile honeypots that could dynamically change their properties to mimic real systems better and evade detection~\cite{dowling2020new-1b1}. Other research focused on using machine learning for post-facto analysis, such as the UNADA system, which applied clustering to NetFlow data from honeypots to autonomously characterize attacks~\cite{bao2015near-f48,dowling2019machine-f43}. While powerful, these systems often relied on aggregated or flow-level data, making them unsuitable for the kind of real-time, first-packet decision-making required for dynamic resource allocation.

\subsection{Evolution of Reinforcement Learning in Honeypot Systems}
Reinforcement Learning (RL) emerged as a particularly promising paradigm for adaptive honeypots because it allows an agent to learn optimal behavior through direct interaction with an environment, a natural fit for the adversarial dynamics of cybersecurity. The core idea is to treat the honeypot as an agent that learns a policy to maximize a reward, which is typically tied to the quality of intelligence gathered.

The evolution of RL in honeypot systems can be traced through several key developments. Pioneering work in this area includes \textbf{RASSH} by Pauna et al., which used Q-learning to enable an SSH honeypot to learn which responses to adversary commands would prolong the session and elicit more interesting behavior~\cite{pauna2014rassh-54e}. This represented a significant step forward, as the honeypot was no longer just a passive recorder but an active participant in the deception. The reward function in RASSH was primarily based on session duration, with longer sessions being rewarded as they were assumed to yield more intelligence.

Subsequent research built upon this foundation. \textbf{Asguard} by Touch and Colin modernized the approach by using a Deep Q-Network (DQN), allowing the agent to handle a much larger state space and learn more nuanced interaction policies~\cite{touch2021asguard-ae5}. Similar to RASSH, Asguard's reward mechanism was also based on session duration, but the DQN architecture enabled more sophisticated decision-making about command responses.

Other researchers have explored using RL to manage the trade-off between interaction and resource consumption. For instance, Veluchamy and Kathavarayan used deep RL to build honeypots that could mitigate DoS attacks~\cite{veluchamy2022deep-98e}, while Limouchi and Mahgoub used RL to optimize thresholds for adapting IoT honeypots~\cite{limouchi2021reinforcement-f6a}.

However, a common thread in this body of work is that the "adaptation" is confined to the behavior of a single, running service (in-service adaptation). The RL agent learns what command to issue or what banner to present. It does not, however, make decisions about the underlying infrastructure itself, such as whether to deploy a honeypot in the first place. Moreover, the reward functions in these systems are typically simplistic, focusing on session duration or resource consumption rather than the quality or novelty of the intelligence gathered.

In contrast, this work proposes a quality-based reward function that values novel and significant adversary interactions over mere session longevity. This represents a considerable advancement over simpler metrics from prior work, addressing the "quality vs. quantity" intelligence problem.

\subsection{Deep Learning for Network Traffic Analysis}
Parallel to the advancements in adaptive honeypots, the broader field of network intrusion detection has been revolutionized by deep learning~\cite{goodfellow2016deep-96f}. Researchers have successfully applied various neural network architectures to the problem of classifying malicious network traffic.

Works like RNN-IDS by Yin et al. demonstrated the effectiveness of Recurrent Neural Networks (RNNs) for capturing the sequential nature of network flows~\cite{yin2017deep-f0b}. Autoencoders have been widely used for unsupervised anomaly detection, with systems like DeepLog~\cite{thuraisingham2017deeplog-22a} and other similar approaches~\cite{shone2018deep-dfe, catillo2022autolog-789, park2018anomaly-3c7} learning a model of normal behavior and flagging deviations. These methods are highly relevant to the "post-interaction log analysis" component of the proposed architecture.

More recently, state-of-the-art research has focused on applying advanced models directly to raw packet data. Sharan et al. demonstrated the use of Transformer models on raw packet bytes for malicious traffic classification, removing the need for manual feature engineering~\cite{sharan2023raw-aab}. While these approaches achieve high classification accuracy, they come with two significant drawbacks for this use case: they are computationally intensive, making them difficult to run in real-time for every single incoming packet, and their primary goal is classification, not the orchestration and resource allocation decision-making that is central to this thesis.

\subsection{Synthesis and Identification of the Research Gap}
The existing body of literature reveals a clear trajectory: from static, passive honeypots to interactive, ML-driven systems. However, it also illuminates a critical, unaddressed gap in the research landscape, which this thesis directly targets. Current research can be broadly categorized into two camps:
\begin{enumerate}
    \item \textbf{In-Service Adaptation:} RL-based honeypots that learn to optimize their interaction \emph{within} a single, already-running service (e.g., choosing the best SSH command response).
    \item \textbf{Offline Classification:} Deep learning models that are highly effective at classifying network traffic or logs as malicious or benign, but are typically too resource-intensive for real-time, per-packet decisions and are not designed to control infrastructure.
\end{enumerate}
The gap lies at the intersection of these two areas: the use of machine learning, specifically reinforcement learning, for the dynamic, real-time \emph{orchestration} of the honeynet infrastructure itself. No existing work focuses on using first-packet analysis to drive an RL agent that makes the foundational decision of whether to expend resources to deploy a containerized, high-interaction honeypot from a pool of available types. This work bridges the divide between single-service adaptation and full infrastructure automation, proposing a system that is adaptive not just in its responses but in its very composition. Specifically, this research introduces the novel application of RL for infrastructure-level orchestration based on real-time analysis of network traffic, a significant advancement over existing approaches that focus solely on in-service adaptation or offline analysis.

\section{Contemporary Threat Landscape Analysis}
\label{sec:threat_landscape}

\subsection{Evolution of Cyber Attack Methodologies}
\noindent
The contemporary cyber threat landscape is characterized by a fundamental shift from opportunistic, single-vector attacks to sophisticated, multi-stage campaigns that leverage advanced techniques and persistent access. This evolution reflects the increasing professionalization of cybercriminal organizations and the growing economic incentives for successful cyber attacks~\cite{enisa2024enisa-658}.

\subsubsection{Advanced Persistent Threats (APTs)}
Advanced Persistent Threats (APTs) represent the most sophisticated category of cyberattacks, typically attributed to nation-state actors or well-resourced criminal organizations~\cite{initiative2011managing-418, agency2025nation-state-74a}. APT campaigns are characterized by their extended duration, refined techniques, and specific strategic objectives. Recent analysis by cybersecurity firms indicates that APT groups are increasingly adopting living-off-the-land techniques, utilizing legitimate system tools and processes to avoid detection by traditional security measures~\cite{crowdstrike2025crowdstrike-b21, verizon20242024-c6e}.

The APT lifecycle typically consists of multiple phases: initial reconnaissance and target selection, initial compromise through spear-phishing or supply chain attacks, establishment of persistent access through backdoors and legitimate credentials, lateral movement to identify and access high-value targets, data exfiltration or system manipulation to achieve strategic objectives, and long-term persistence to maintain access for future operations~\cite{hutchins2011intelligence-driven-c04}. Each phase presents unique challenges for detection and mitigation, requiring adaptive defense mechanisms that can identify subtle indicators across extended periods. ~\footnote{Subtle indicators may include low-and-slow scanning, use of encrypted C2 channels, or living-off-the-land binaries (LoLBins) that are difficult to distinguish from legitimate administrative activity.}

Notable APT groups, such as APT29 (Cozy Bear), APT28 (Fancy Bear), and the Lazarus Group, have demonstrated remarkable adaptability in their techniques, continuously evolving their tools and methods in response to defensive measures~\cite{crowdstrike2025crowdstrike-b21}. This adaptability highlights the need for equally adaptable defensive technologies that can grow alongside the capabilities of threat actors.

\subsubsection{Ransomware Evolution and Industrialization}
The ransomware threat landscape has undergone a dramatic transformation, evolving from simple file encryption attacks to complex, multi-stage operations that combine data theft, system disruption, and extortion. Modern ransomware operations often follow a double or triple extortion model, where adversaries not only encrypt the victim's data but also threaten to publicly release sensitive information and target the victim's customers, partners, or stakeholders~\cite{chainanalysis20252025-1cc}.

Ransomware-as-a-Service (RaaS) platforms have democratized access to sophisticated attack capabilities, enabling less technically skilled actors to launch devastating attacks. Primary RaaS operations such as Conti, REvil, and LockBit have operated with corporate-like structures, including customer support, affiliate programs, and quality assurance processes~\cite{chainanalysis20252025-1cc, europol2025internet-692}. The industrialization of ransomware has led to an increase in attack frequency, sophistication, and economic impact.

\begin{figure*}[htbp]
\centering
\includegraphics[width=0.75\textwidth]{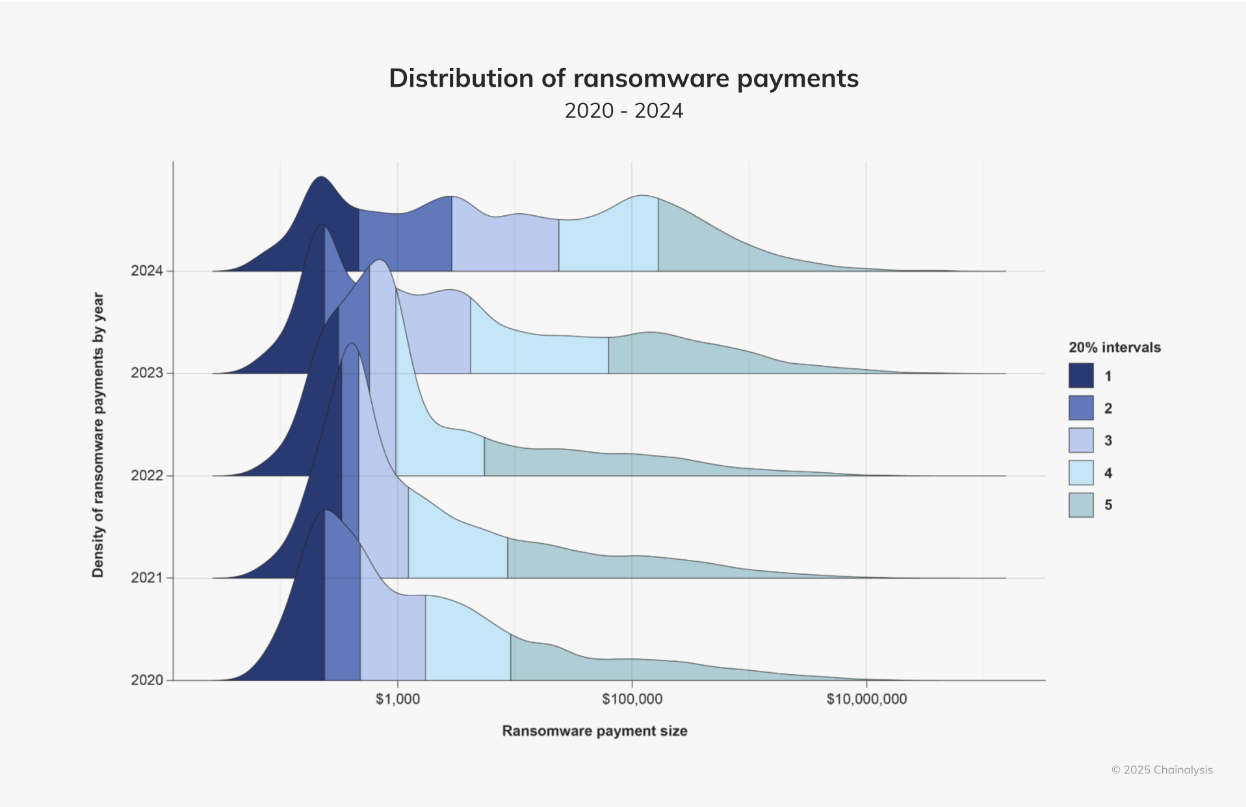}
\caption{Ransomware Payment Size Analysis by Chainanalysis.~\cite{chainanalysis20252025-1cc}}
\label{fig:ransomware-payment}
\end{figure*}

The emergence of targeted ransomware attacks against critical infrastructure has significantly raised the stakes. Attacks on Colonial Pipeline~\cite{agency2021darkside-e3a}, JBS~\cite{securityscorecard2021jbs-127}, and Kaseya~\cite{nguyen-duy2021new-d0c} demonstrated the potential for ransomware to disrupt essential services and supply chains, highlighting the urgent need for more effective defensive technologies. These attacks often involve extensive reconnaissance, lateral movement, and privilege escalation before the final ransomware deployment, creating multiple opportunities for detection by adaptive honeynets. As illustrated in Figure~\ref{fig:ransomware-payment}, the economic scale of these attacks is substantial.

\subsubsection{Supply Chain Attacks and Third-Party Risks}
Supply chain attacks have emerged as one of the most concerning threat vectors, as they leverage the trust relationships between organizations and their technology providers. High-profile incidents such as SolarWinds~\cite{fireeye2020highly-c3d}, Codecov~\cite{codecov2021post-mortem-caa}, XZ~\cite{informationstechnik2024kritische-5a1} and Kaseya~\cite{nguyen-duy2021new-d0c} have demonstrated how adversaries can compromise software development and distribution processes to gain access to thousands of downstream victims simultaneously~\cite{mandiant2025m-trends-03a}.

These attacks are particularly challenging to detect because they often involve legitimate software and trusted communication channels. Traditional security measures may not identify malicious activity that appears to originate from trusted sources. The complexity of modern software supply chains, with multiple layers of dependencies and third-party components, creates numerous potential attack vectors that are difficult to monitor comprehensively~\cite{verizon20242024-c6e}.

The increasing adoption of cloud services and software-as-a-service (SaaS) platforms has expanded the attack surface for supply chain compromises. Adversaries can potentially gain access to multiple organizations by compromising a single cloud service provider or SaaS platform. This interconnectedness requires security solutions that can identify subtle anomalies in trusted communications and detect indicators of compromise across complex, distributed environments~\cite{enisa2024enisa-658}.

\subsection{Threat Actor Categorization and Capabilities}
\noindent
Understanding the diverse landscape of threat actors is essential for developing effective defensive strategies. Different categories of threat actors possess varying capabilities, motivations, and operational patterns that influence their attack methodologies and the types of threats they pose~\cite{johnsonnoyearguide-761}.

\subsubsection{Nation-State Actors}
Nation-state threat actors represent the most sophisticated and well-resourced category of cyber adversaries. These actors typically have access to zero-day exploits, custom malware, and extensive infrastructure for conducting long-term campaigns. Their objectives often include espionage, intellectual property theft, disrupting critical infrastructure, and conducting influence operations~\cite{crowdstrike2025crowdstrike-b21}.

Nation-state actors are characterized by their patience and persistence, often maintaining access to target networks for years while carefully avoiding detection. They employ sophisticated operational security practices, including the use of legitimate infrastructure, encrypted communications, and careful timing of activities to blend with regular network traffic. Their advanced capabilities and resources make them particularly challenging opponents for traditional security measures~\cite{agency2025nation-state-74a, agency2024prc-ad0}.

Recent trends indicate that nation-state actors are increasingly targeting cloud infrastructure, software supply chains, and managed service providers to maximize their reach and impact. They are also becoming more aggressive in their operations, with some groups conducting destructive attacks in conjunction with traditional espionage activities~\cite{mandiant2025m-trends-03a}.

\subsubsection{Cybercriminal Organizations}
Professional cybercriminal organizations have evolved into sophisticated enterprises with specialized roles, hierarchical structures, and business-like operations. These organizations often focus on financially motivated attacks, including ransomware, banking fraud, cryptocurrency theft, and data monetization~\cite{europol2025internet-692}.

Modern cybercriminal groups demonstrate remarkable operational sophistication, employing techniques such as initial access brokers who specialize in gaining entry to target networks, ransomware operators who deploy and manage encryption attacks, money laundering services that facilitate the conversion of stolen funds, and customer support operations that assist victims with ransom payments~\cite{europol2025internet-692}.

The professionalization of cybercrime has led to increased specialization and efficiency in attack operations. Criminal groups often purchase access to compromised networks from initial access brokers, reducing the time and resources required to launch attacks. This ecosystem approach enables the rapid scaling of criminal operations, making it more difficult for law enforcement to disrupt entire criminal enterprises~\cite{verizon20242024-c6e}.

\subsubsection{Hacktivist Groups}
Hacktivist organizations are motivated by political, social, or ideological objectives rather than financial gain. These groups often target organizations or governments that they perceive as opposing their beliefs or values. While hacktivist attacks may be less sophisticated than nation-state operations, they can still cause significant disruption and reputational damage~\cite{bundeskriminalamt2016hacktivists-8eb}.

Hacktivist groups often employ tactics such as distributed denial-of-service (DDoS) attacks, website defacements, data leaks, and social media manipulation~\cite{bundeskriminalamt2016hacktivists-8eb}. Recent trends indicate that some hacktivist groups are adopting more sophisticated techniques, including ransomware and destructive attacks, blurring the lines between hacktivism and cybercrime~\cite{enisa2024enisa-658}.

The decentralized nature of many hacktivist organizations makes them difficult to predict and counter. These groups often operate through loose networks of volunteers with varying skill levels, making their attack patterns less predictable than those of professional criminal organizations or nation-state actors.

\subsubsection{Insider Threats}
Insider threats represent a unique category of risk that combines legitimate access with malicious intent or negligent behavior. Malicious insiders may be motivated by financial gain, revenge, ideology, or coercion by external actors. Negligent insiders may inadvertently create security vulnerabilities through poor security practices or by falling victim to social engineering~\cite{cappelli2012cert-b03}.

Insider threats are particularly challenging to detect because they often involve legitimate access to systems and data. Traditional perimeter-based security measures are ineffective against threats that originate from within the organization~\cite{cappelli2012cert-b03}. Detecting insider threats requires behavioral analysis, anomaly detection, and careful monitoring of privileged access activities~\cite{verizon20242024-c6e}.

Recent trends indicate that insider threats are becoming more sophisticated, with some malicious insiders collaborating with external threat actors to maximize the impact of their activities. The increasing adoption of remote work and cloud services has expanded the potential attack surface for insider threats, making comprehensive monitoring more challenging.

\subsection{Attack Vector Analysis}
\noindent
Modern cyberattacks employ diverse attack vectors that exploit different aspects of an organization's infrastructure, processes, and human behavior. Understanding these attack vectors is essential for developing comprehensive defensive strategies~\cite{scarfonenoyearguide-6ed}.

\subsubsection{Network-Based Attack Vectors}
Network-based attacks remain a primary concern for organizations, as they can provide adversaries with initial access to target environments and enable lateral movement within compromised networks. Common network attack vectors include exploitation of unpatched vulnerabilities in network services, brute force attacks against authentication systems, man-in-the-middle attacks against network communications, and exploitation of misconfigured network devices and services.~\cite{verizon20242024-c6e, noauthornoyearmitre-518, enisa2024enisa-658}

The increasing adoption of cloud services and remote work has significantly expanded the network attack surface. Organizations now must secure not only their traditional on-premises infrastructure but also cloud environments, remote access solutions, and the network connections between these diverse components. This expanded attack surface creates new opportunities for adversaries and new challenges for defenders.~\cite{cloudsecurityalliance2024top-589, enisa2024enisa-658}

Network segmentation and zero-trust architectures have emerged as critical defensive strategies, but they require sophisticated monitoring and analysis capabilities to be effective~\cite{rose2020zero-8e6}. Adaptive honeynets can play a crucial role in network defense by providing early warning of attack activities and detailed intelligence about adversary techniques and objectives.

\subsubsection{Application-Level Attack Vectors}
Application vulnerabilities continue to be a significant source of security incidents, with web applications being highly desirable targets due to their accessibility and the valuable data they often contain. Common application attack vectors include SQL injection, cross-site scripting (XSS), remote code execution, authentication bypass, and business logic flaws~\cite{owasp2021owasp-c95}.

The rapid pace of software development and deployment in modern organizations often leads to security vulnerabilities being introduced and deployed to production environments. DevSecOps practices aim to address this challenge by integrating security testing into the development pipeline~\cite{myrbakken2017devsecops-48f}. Still, the complexity of modern applications and the pressure for rapid deployment continue to create opportunities for adversaries.

API security has become increasingly important as organizations adopt microservices architectures and API-driven integrations. APIs often lack the same level of security controls as traditional web applications, creating new avenues for sophisticated threat actors to exploit~\cite{owasp2023owasp-6ac}. The proliferation of APIs and the complexity of API ecosystems make comprehensive security monitoring a challenging task.

\subsubsection{Social Engineering and Human-Centric Attacks}
Social engineering attacks exploit human psychology and behavior rather than technical vulnerabilities, making them particularly effective and challenging to defend against~\cite{mitnick2003art-e2d}. These attacks often serve as the initial vector for more sophisticated campaigns, providing adversaries with legitimate credentials and access to target environments.

Phishing attacks have evolved significantly in sophistication, with adversaries employing detailed reconnaissance to create highly targeted and convincing messages~\cite{alkhalil2021phishing-7e1}. Spear-phishing attacks often incorporate information gathered from social media, corporate websites, and data breaches to create personalized messages that are difficult to distinguish from legitimate communications.

Business email compromise (BEC) attacks represent a particularly damaging category of social engineering, often resulting in significant financial losses. These attacks typically involve compromising or spoofing executive email accounts to authorize fraudulent financial transactions. The success of BEC attacks demonstrates the importance of implementing comprehensive security awareness training and technical controls that can detect anomalous communication patterns.~\cite{investigation2024internet-1d1, ogwo-ude2023business-6dc}

\subsection{Emerging Threat Trends}
\noindent
The cyber threat landscape continues to evolve rapidly, with new trends and techniques emerging regularly. Understanding these emerging trends is essential for developing adaptive defensive strategies that can address future threats~\cite{manshaei2013game-b0b}.

\subsubsection{AI-Powered Attacks}
The increasing availability of artificial intelligence and machine learning technologies is enabling adversaries to develop more sophisticated and automated attack capabilities. AI-powered attacks can include automated vulnerability discovery, intelligent evasion of security controls, deepfake technology for social engineering, and automated spear-phishing campaigns~\cite{brundage2018malicious-320}.

Machine learning algorithms can be used to analyze defender behavior and adapt attack strategies in real-time, creating a dynamic adversarial environment where both adversaries and defenders employ AI technologies. This trend underscores the importance of developing AI-powered defensive technologies, such as adaptive honeypots, that can effectively counter AI-powered attacks.

The democratization of AI technologies through cloud services and open-source frameworks is making these capabilities accessible to a broader range of threat actors, potentially accelerating the adoption of AI-powered attack techniques across different threat actor categories~\cite{enisa2023enisa-e40}.

\subsubsection{Cloud-Native Attacks}
As organizations increasingly adopt cloud infrastructure and cloud-native technologies, adversaries are developing specialized techniques for targeting these environments. Cloud-native attacks often exploit misconfigurations in cloud services, abuse legitimate cloud APIs, and leverage the shared responsibility model to target areas where security controls may be unclear or inadequate.~\cite{cloudsecurityalliance2024top-589}

Container and Kubernetes security has become a particular concern as organizations adopt containerized applications and orchestration platforms. Adversaries are developing techniques to escape container environments, exploit Kubernetes misconfigurations, a threat detailed in hardening guides~\cite{cybersecurity2022kubernetes-176}, and abuse container registries to distribute malicious images~\cite{noauthornoyearkubernetes-2f9}.

The complexity of cloud environments and the rapid pace of cloud service development create ongoing challenges for security teams. Traditional security tools and processes may not be effective in cloud environments, requiring new approaches to monitoring, detection, and response.

\subsubsection{IoT and Edge Computing Threats}
The proliferation of Internet of Things (IoT) devices and edge computing infrastructure is creating new attack vectors and expanding the overall attack surface. IoT devices often lack robust security controls and may be difficult to update or patch, making them attractive targets for adversaries~\cite{neshenko2019demystifying-447}.

IoT botnets have become a significant threat, with adversaries compromising large numbers of IoT devices to conduct distributed denial-of-service attacks, cryptocurrency mining, and other malicious activities~\cite{antonakakis2017understanding-14e}. The scale and distributed nature of IoT botnets make them challenging to detect and mitigate using traditional security approaches.

Edge computing environments present unique security challenges due to their distributed nature and the need to balance security with performance and latency requirements. Securing edge environments requires new approaches to monitoring and threat detection that can operate effectively in resource-constrained and distributed environments~\cite{xiao2019edge-70e}.

\subsection{Impact Assessment and Economic Consequences}
\noindent
The economic impact of cyberattacks continues to grow, with organizations facing direct financial losses, operational disruptions, regulatory penalties, and long-term reputational damage. Understanding these impacts is essential for justifying investments in advanced defensive technologies such as adaptive honeynets.

\subsubsection{Direct Financial Losses}
Cyber attacks can result in immediate financial losses through theft of funds, ransom payments, system recovery costs, and business interruption. The average cost of a data breach reached \$4.44 million in 2025, according to IBM's Cost of a Data Breach Report, with costs varying significantly based on the size of the organization, industry sector, and geographic location~\cite{ibm2025cost-01a}.

Exceptionally high direct costs are imposed by ransomware attacks, forcing difficult decisions on whether ransoms should be paid, system restoration should be pursued, or prolonged operational downtime should be accepted. In 2024 alone, ransomware payments were reported to total \$813.55 million~\cite{chainanalysis20252025-1cc}.

The increasing sophistication of financial fraud attacks, including business email compromise (BEC) and payment fraud, has resulted in significant direct economic losses for organizations. The FBI's Internet Crime Complaint Center reported over \$2.77 billion in losses from business email compromise attacks in 2024 alone~\cite{investigation2024internet-1d1}.

\subsubsection{Operational Disruption and Recovery Costs}
Beyond direct financial theft, cyberattacks often result in significant operational disruptions that can have cascading effects throughout an organization and its supply chain. System downtime, data recovery efforts, and the need to implement alternative processes can result in substantial indirect costs that may exceed the direct financial impact of an attack.~\cite{ibm2025cost-01a}

The time required to recover from a sophisticated cyber attack fully can extend for months or even years, particularly for attacks that involve extensive data theft or system compromise. During this recovery period, organizations may face reduced operational efficiency, increased security costs, and ongoing uncertainty about the full scope of the compromise.~\cite{ibm2025cost-01a}

Critical infrastructure attacks can have particularly severe operational impacts, potentially affecting essential services such as power generation, transportation, healthcare, and communications. The 2021 Colonial Pipeline ransomware attack demonstrated how cyberattacks on critical infrastructure can have widespread economic and social impacts that extend far beyond the directly targeted organization.~\cite{ferguson2021fbi-a05}

\subsubsection{Regulatory and Legal Consequences}
The increasing focus on data protection and cybersecurity regulations has created significant potential for legal and regulatory consequences following cyberattacks. Organizations may face substantial fines under rules such as the General Data Protection Regulation (GDPR)~\cite{parliament2016regulation-bc4} and the California Consumer Privacy Act (CCPA)~\cite{california2018assembly-985}, as well as various industry-specific requirements.

Beyond regulatory fines, organizations may face civil litigation from customers, partners, and shareholders who are affected by cyber attacks. Class action lawsuits following major data breaches can result in settlement costs that reach hundreds of millions of dollars, particularly when sensitive personal or financial information is compromised.~\cite{comission2024ftc-057, comission2024equifax-133}

\section{The Art of Invisibility: Evading Honeynet Detection}
\label{sec:evasion}

\subsection{Introduction: The Adversary's Advantage}
A honeynet's value is directly proportional to its believability. Adversaries, aware of deception technologies, are strongly motivated to detect them to avoid analysis, prevent wasting resources, and, in some cases, to study the defenders' techniques. This leads to a cat-and-mouse game where adversaries actively ``fingerprint'' systems to expose the artifacts of a controlled environment. If a honeynet can be easily identified, its ability to capture high-fidelity intelligence from sophisticated actors is severely reduced. ~\footnote{High-fidelity intelligence implies detailed, actionable data about adversary tools, infrastructure, and intent, as opposed to the low-fidelity data from automated scans.}

\subsection{State-of-the-Art in Honeynet Detection (The ``Offense'')}
Over the years, adversaries have developed a sophisticated arsenal of techniques to unmask honeynets. These methods, as surveyed by Holz et al.~\cite{holz2005detecting-508}, primarily fall into three categories:

\subsubsection{Environmental Artifacts}
The most common detection vector is the search for artifacts left by virtualization or containerization~\cite{rhling2019standardized-2b8, papazis2019detecting-0f7}. In their foundational work on detecting virtual machines, subtle differences in CPU instructions, memory layout, and I/O timing can betray the presence of a hypervisor. Adversaries search for specific MAC address prefixes (e.g., `00:0C:29` for VMware), unique hardware identifiers, or the presence of guest tools drivers that would not exist on a bare-metal machine.

\subsubsection{Behavioral Quirks and Interaction Probes}
Low-interaction honeypots, while efficient, often exhibit behavioral quirks that can be detected. They may present service banners that are slightly different from a real service, have unnatural response delays, or implement only a partial set of a protocol's features. Sophisticated adversaries employ multi-stage fingerprinting frameworks~\cite{srinivasa2023gotta-f3e}, which automate a series of interaction probes. These frameworks test for known vulnerabilities, measure system responses, and analyze protocol handshakes to determine if a target is a honeypot. Another critical vulnerability is \textbf{handoff inconsistency}, which can occur in multi-layered systems like ADLAH. When a connection is passed from a low-interaction sensor to a high-interaction honeypot, any discrepancy in the protocol handshake (e.g., different SSH server banners, cipher suites, or authentication methods) can reveal the deception.

\subsubsection{Network-Level Indicators}
Analysis of network traffic can also reveal a honeynet. For example, Time-To-Live (TTL) values in IP packets can sometimes suggest that traffic has passed through an additional hop (the hypervisor) compared to a real host on the same subnet. Furthermore, TCP/IP stack fingerprinting tools like `nmap` can identify discrepancies between the advertised operating system and the actual network behavior. A significant temporal artifact is \textbf{deployment latency}; the delay introduced when provisioning a honeypot on demand can create a noticeable pause during the connection handshake, which is highly suspicious and can be used for fingerprinting.

\subsection{State-of-the-Art in Counter-Evasion (The ``Defense'')}
In response, defenders have developed techniques to make their honeynets more credible:

\subsubsection{High-Interaction Honeypots}
The most effective defense is to use a high-interaction honeypot that provides a real, full-fledged operating system and applications. This minimizes behavioral quirks, as the adversary is interacting with the genuine software. Systems like Argos~\cite{portokalidis2006argos-b24} pushed this concept further by using dynamic taint analysis to provide high-interaction emulation with better performance and control, making them much more challenging to detect than their low-interaction counterparts.

\subsubsection{System Hardening and Obfuscation}
This involves manually or semi-automatically altering the honeypot's environment to remove noticeable artifacts. This can include changing MAC addresses, modifying the registry on Windows systems, and eliminating vendor-specific drivers or files associated with the virtualization platform. However, this is a painstaking process that must be constantly updated as new detection techniques emerge.

\needspace{5\baselineskip}
\subsection{ADLAH's Approach to Advanced Stealth}
ADLAH is designed to move beyond these static defenses by treating stealth as a dynamic, intelligent process.

\subsubsection{Current Capabilities: Dynamic Reconfiguration}
The current ADLAH prototype leverages its container orchestration backbone (k3s) to implement a ``moving target'' defense. By rapidly deploying and destroying honeypot instances from a pool of different container images (e.g., various Linux distributions, different service versions), it presents a constantly changing attack surface. This dynamic reconfiguration complicates long-term fingerprinting, as an adversary cannot rely on a single set of artifacts to identify the honeynet over time.

\subsubsection{Target Architecture: Intelligent, Adaptive Stealth (Vision)}
The full vision for ADLAH integrates stealth directly into the reinforcement learning agent's decision-making process. The goal is not merely to engage an adversary, but to engage without being discovered. This is achieved through two key concepts:
\begin{itemize}
    \item \textbf{RL-Driven Obfuscation:} The RL agent's reward function will be modified to explicitly penalize the deployment of honeypot configurations that are known to be easily fingerprinted. For instance, if interactions with a specific honeypot type consistently lead to behavior indicative of detection (e.g., the use of fingerprinting tools), the agent will learn to avoid that configuration. A high detectability score should result in a \textbf{negative reward}, teaching the agent to avoid configurations that can be easily fingerprinted.
    \item \textbf{Observability and Feedback Loop:} The system will be enhanced to monitor for signs of detection actively. By correlating adversary actions with known fingerprinting techniques, ADLAH can create a ``detectability score'' in real-time. This score becomes a critical input for the RL agent, allowing it to adapt its strategy dynamically, perhaps by migrating the interaction to a higher-interaction, more credible honeypot if it suspects it is being probed. This creates a closed-loop system where the honeynet learns to become progressively stealthier over time.
\end{itemize}

\section{Full Adaptive Honeynet Architecture}
\label{sec:architecture}

\begin{figure*}[!t] 
    \centering
    \includegraphics[width=\textwidth]{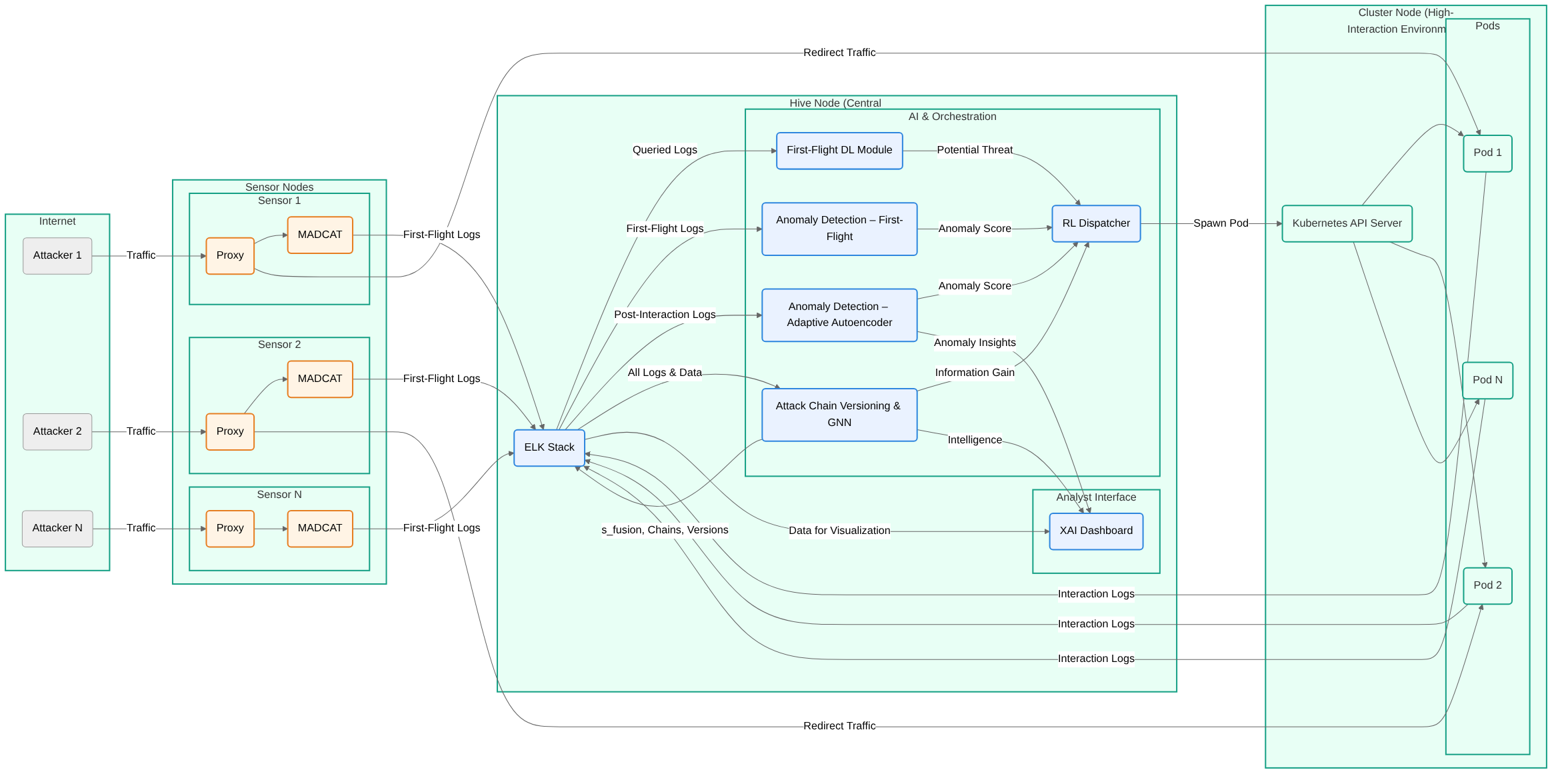}
    \caption{Visual overview of the adaptive honeynet architecture.}
    \label{fig:arch_overview}
\end{figure*}
\begin{figure}[htbp]
    \centering
    \includegraphics[width=\linewidth]{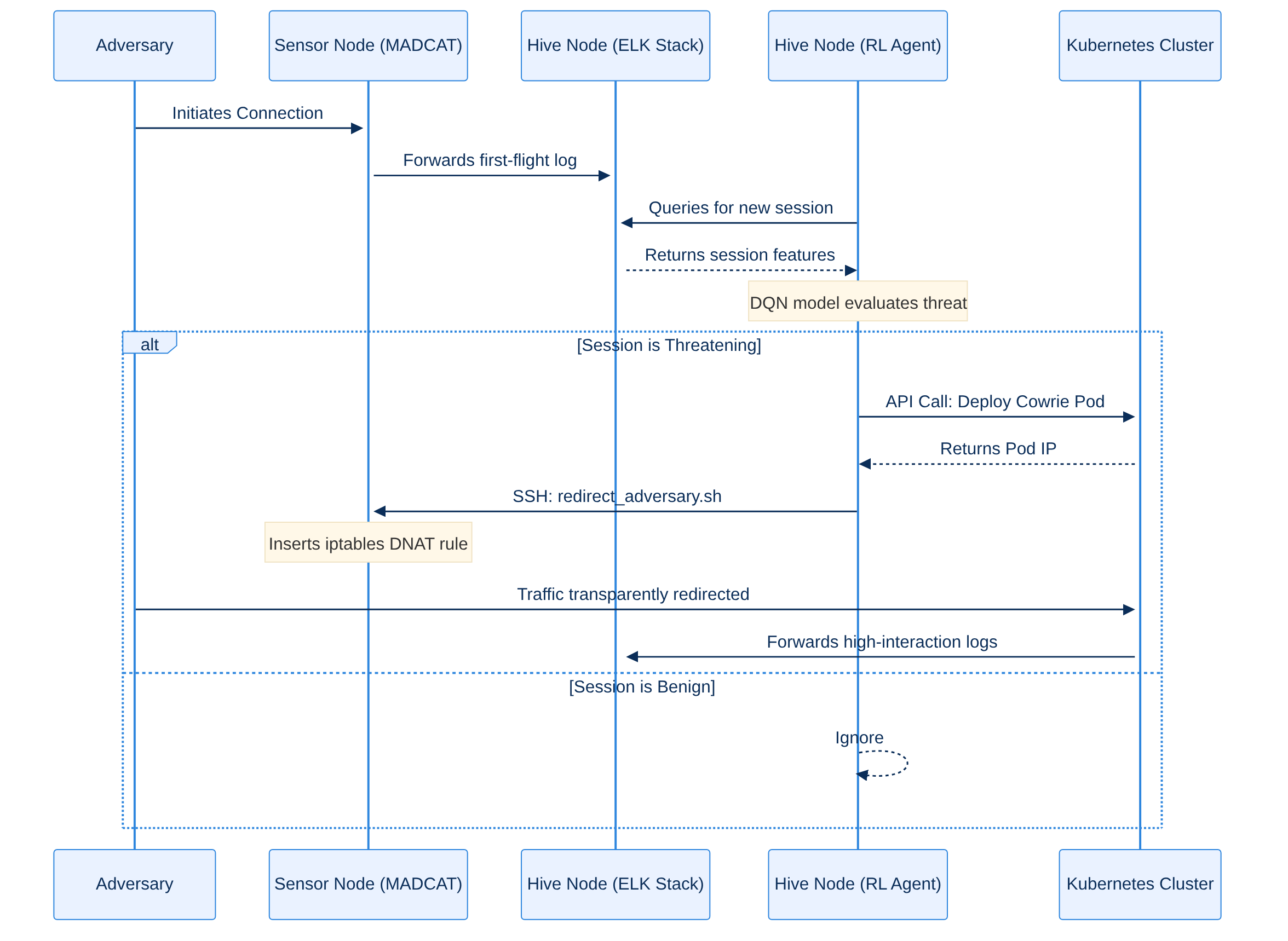}
    \caption{Detailed data flow of an adversary session, from initial contact to high-interaction engagement.}
    \label{fig:data_flow_detailed}
\end{figure}

The proposed adaptive honeynet architecture, illustrated in Figure~\ref{fig:arch_overview} and with its data flow detailed in Figure~\ref{fig:data_flow_detailed}, is a modular, cloud-native system designed for advanced threat analysis, dynamic deception, and real-time analyst support. It integrates four key conceptual components: a low-interaction Sensor Node for initial contact, a central "Hive" for data aggregation and decision-making, a lightweight Kubernetes (k3s) cluster for dynamic high-interaction honeypot deployment, and a sophisticated AI pipeline for threat analysis and orchestration. This section describes the high-level design and data flow of this architecture, while Section~\ref{sec:prototype} will detail the specific implementation of the working prototype.

\subsection{First-Flight Analysis (Prototype) \& Reinforcement Learning Deployment}
The prototype makes deployment decisions from live telemetry indexed in Elasticsearch rather than exclusively from raw first-packet bytes. An event source polls Elasticsearch for recent documents and forms short, per-source-IP sequences that are fed to a reinforcement learning agent. At each step, the agent observes a state vector derived from the most recent events of a single source IP, chooses an action (\texttt{deploy} or \texttt{wait}), and later receives a delayed reward based on interactions of the adversary with the honeypot pod. When a deployment is triggered, the system configures connection tracking on the Sensor Node, transparently forwarding subsequent traffic from the attacker to the designated honeypot pod. Although the broader vision includes multiple honeypot types, the current action space is binary and focused on deciding if and when to escalate a source to a high-interaction pod.

\subsection{Post-Interaction Log Analysis and Anomaly Detection}
All logs from both low- and high-interaction sources are centrally aggregated in the Hive. While the prototype's RL agent acts on first-packet data, the whole architecture envisions a sophisticated \textbf{AI Analytics Pipeline} for post-interaction analysis of logs from high-interaction honeypots like Cowrie. This pipeline serves as a critical component for identifying novel adversary behaviors that deviate from previously observed patterns and for continuously improving the system's understanding of threat landscapes.

\subsubsection{Adaptive Autoencoder for Log Anomaly Detection}
A core component of this pipeline is an \textbf{adaptive autoencoder}, designed to analyze sequences of events from high-interaction logs. Unlike traditional anomaly detection systems that are trained once on a static dataset, this autoencoder is designed for continuous online learning. The intelligence refinement loop leverages both the autoencoder approach for novelty detection and complementary techniques from the literature, such as RNNs/LSTMs for TTP classification, to provide a comprehensive analysis of adversary behavior.

The process is as follows:
\begin{enumerate}
    \item \textbf{Feature Extraction:} Logs from a completed high-interaction session (e.g., a series of commands entered in Cowrie) are transformed into a numerical feature vector. This can include features such as command frequency, use of specific tools, session duration, and payload characteristics.
    \item \textbf{Anomaly Scoring:} The feature vector $\mathbf{Y} \in \mathbb{R}^d$ is fed into the autoencoder, producing a reconstruction $\hat{\mathbf{x}}$. The anomaly score is computed as the mean squared reconstruction error:
    \begin{equation}
    s(\mathbf{x}) = \frac{1}{d} \sum_{j=1}^d \left( x_j - y_j \right)^2,
    \label{eq:mse_score}
    \end{equation}
    where $\mathbf{x}$ is the original feature vector, $\mathbf{y}$ is its reconstruction, and $d$ is the number of features.
    This formulation follows established practice in autoencoder-based intrusion detection~\cite{shone2018deep-dfe}.
    A higher $s(\mathbf{x})$ indicates that the observed behavior deviates more strongly from what the model has learned as ``normal.''
    \item \textbf{Online Training:} The autoencoder is continuously retrained on feature vectors from incoming logs, allowing its definition of ``normal'' to evolve over time. As specific attack patterns become frequent, the model learns to reconstruct them with low error, so that a high reconstruction error becomes a strong indicator of \emph{truly novel} or significantly mutated attack patterns. This mitigates the impact of concept drift~\cite{gama2014survey-898, lu2018learning-4db}, where adversary tactics change over time.
\end{enumerate}

This mechanism allows the system to automatically surface the most interesting and unusual interactions for security analysts, prioritizing novel threats over the background noise of common automated attacks. If this approach is working, needs to be evaluated. 

\subsubsection{Integration with Reinforcement Learning Reward}
The anomaly score generated by the adaptive autoencoder provides a powerful signal for refining the RL agent's reward function. The long-term vision is to move beyond a purely quantity-based reward to a quality-based one, addressing the "quality vs. quantity" intelligence problem. The reward function could be augmented as follows:

\begin{equation}
    R = \big( 1 + \omega \cdot \tilde{A}_{\text{score}} \big) \cdot N_{\text{logs}} - \lambda \cdot C,
    \label{eq:anomaly_weighted_reward}
\end{equation}

where $A_{\text{score}}$ is the normalized reconstruction error from the autoencoder for that session's logs, and $\omega$ is a weighting hyperparameter. This creates a system that not only learns to engage adversaries but preferentially learns to engage \emph{novel} and \emph{interesting} adversaries, thereby maximizing the value of the collected intelligence.

\subsection{Attack Chain Detection}
A central innovation of the architecture is the reconstruction of attack chains. All logs from a single source Internet Protocol (IP) are grouped as a candidate attack chain. To address distributed or proxy-based attacks, an advanced AI correlation model analyzes behavioral indicators (e.g., command sequences, timing, tool fingerprints) to link sessions from different IPs that likely originate from the same adversary. This enables the system to track complex, multi-session attack campaigns and attribute activity to persistent adversaries.

The attack chain correlation engine employs graph-based algorithms to model relationships between sessions based on temporal proximity, behavioral similarity, and shared infrastructure~\cite{zhou2020graph-a44}. Session features are embedded in a high-dimensional space where similarity metrics capture both explicit feature matching and implicit behavioral patterns. Machine learning classifiers categorize detected chains based on attack type, complexity, and automation level, enabling a prioritized response to high-risk campaigns.

Temporal analysis techniques identify patterns in attack timing that may indicate the use of automated tools, human-operated attacks, or coordinated campaigns across multiple threat actors. The system maintains persistent adversary profiles that evolve, capturing changes in tactics, techniques, and procedures (TTPs) as adversaries adapt to defensive measures. Cross-organizational correlation capabilities enable the system to identify attack campaigns that span multiple organizations, providing valuable intelligence about broader threat actor activities.

\subsection{Bot Behavior Versioning \& MITRE Mapping}
The system is designed to identify and label recurring automated bot activity. Detected attack chains are fingerprinted using behavioral signatures that capture both static and dynamic characteristics of the attack tools. These signatures are based on command sequences, timing patterns, protocol usage, and other behavioral indicators that remain consistent across different deployments of the same bot family.

The fingerprinting process employs fuzzy matching techniques to identify variants of known bot families, enabling the system to track the evolution of automated attack tools over time. Detected attack chains are mapped to known Tactics, Techniques, and Procedures (TTPs) from the MITRE ATT\&CK framework~\cite{noauthornoyearmitre-518}, providing standardized categorization that facilitates threat intelligence sharing, often using standards like STIX™~\cite{open2021stix-824, kent2006guide-703}, and correlation with external security tools. The mapping process uses semantic similarity measures to associate observed behaviors with the most relevant ATT\&CK techniques, even when the exact implementation differs from documented examples.

If a new attack chain closely matches a known bot pattern but exhibits subtle changes, it is flagged as a new variant and versioned accordingly using a hierarchical versioning scheme. Versioning tracks changes in payload structures, command sequences, evasion techniques, and communication protocols, providing detailed lineage information about how bot families evolve in response to defensive measures. These insights are tracked and managed through a dedicated GUI module, which supports longitudinal bot tracking and threat intelligence enrichment.

\subsection{Adaptive High-Interaction Pods}
Unlike static honeypots, the architecture supports adaptive pod behavior. Deployed pods are capable of modifying their responses in real-time, aiming to maximize information gain about adversary tools and behavior. This may include dynamic response crafting, interactive deception, or runtime behavioral modulation based on observed adversary input, enabling deeper engagement with both automated and human-driven threats.

The adaptive pod framework utilizes a plugin-based architecture that enables the dynamic loading of various response modules based on the detected threat profile. Response modules are designed to mimic specific vulnerable services while incorporating subtle variations that can help identify sophisticated adversaries who may be probing for inconsistencies. The system maintains a library of response templates that can be combined and modified in real-time to create convincing deceptions.

Interactive deception techniques include progressive disclosure of system information, where adversaries are gradually provided with more detailed system information as they demonstrate persistence and capability. This approach helps distinguish between opportunistic scanners and targeted adversaries while maximizing the intelligence value of extended engagements. Runtime behavioral modulation adjusts pod responses based on real-time analysis of adversary behavior, ensuring that interactions remain credible while collecting maximum intelligence about adversary tools and methodologies.

\needspace{4\baselineskip}
\subsubsection{Deep Emulation Portfolio (Integrable Pods)}
After escalation and pod provisioning, the orchestrator selects which high-interaction backend to instantiate. In addition to Cowrie~\cite{teamnoyearcowrie-975}, the following projects can be integrated as standalone pods:
\begin{itemize}
  \item Conpot (ICS/SCADA)~\cite{mushorgnoyearconpot-ea9}  industrial/OT protocols.
  \item Dionaea (malware catcher)~\cite{dinotoolsnoyeardinoaea-183}  collects shellcode and malware samples across common services.
  \item Glastopf (web)~\cite{mushorgnoyearglastopf-080}  web applications/HTTP attacks.
  \item Heralding (credential catcher)~\cite{johnnykv2025heralding-099}  targeted capture of authentication attempts.
  \item Honeytrap (framework)~\cite{honeytrap2025notitle-c67}  protocol-agnostic honeypot framework.
  \item Elastichoney~\cite{wright2025elastichoney-09b}  Elasticsearch-Decoys.
  \item Wordpot~\cite{gbrindisi2025wordpot-385}  WordPress-specific decoys.
  \item IoTPOT (reference design)~\cite{pa2015iotpot-e21}  IoT/Telnet campaigns.
  \item Amun~\cite{zeroq2025amun-982}  complementary legacy environments.
\end{itemize}

\needspace{4\baselineskip}
\subsubsection{Adaptive Honeypots}
In addition to standard pods, research-grade adaptive backends can be integrated (e.g., Asguard~\cite{touch2021asguard-ae5}, Self-Guarded~\cite{touch2022comparison-af3}, QRASSH/RASSH~\cite{pauna2019rewards-e5e, pauna2018qrassh-db1, pauna2014rassh-54e}, HoneyIoT~\cite{boureanu2023honeyiot-893}). These serve as building blocks for RL- or rule-based adaptation of service interaction and stealth strategies.

\subsubsection{Attacker-Dedicated Emulation Networks}
Future iterations of the architecture need not constrain adversary engagement to a single high-interaction pod. 
Instead, each escalated adversary could be provisioned with an isolated, attacker-dedicated emulation network that more closely mirrors a real enterprise environment. 
Such a network could contain multiple interconnected pods simulating heterogeneous hosts and services, allowing the adversary to perform privilege escalation, lateral movement, and multi-stage attacks within a contained virtual environment. 
For example, an attacker might encounter a simulated Active Directory domain, pivot between workstations and servers, or exploit vulnerable services on internal segments. 
This richer engagement surface would enable the collection of higher-fidelity behavioral data, particularly regarding post-compromise tactics, techniques, and procedures (TTPs), and support more realistic testing of adaptive defense and deception strategies.

\subsection{Visualized Attack Chains \& Analyst Interface}
A graphical user interface (GUI), which can be built with tools like Kibana~\cite{noauthornoyearkibana-ea0} or Grafana~\cite{noauthornoyeargrafana-42b, noauthornoyeargrafana-42b}, provides interactive visualizations of attack chains, session flows, and anomaly scores, often powered by backends like Prometheus~\cite{noauthornoyearprometheus-7ee, rabenstein2015prometheus-e27} and OpenTelemetry~\cite{noauthornoyearopentelemetry-84f}. Analysts can manage bot version history, track MITRE ATT\&CK correlations, and review AI-based chain linking. The interface also offers real-time insight into deployed pods and the RL agent's decision process, supporting both operational monitoring and forensic investigation.

The visualization framework utilizes force-directed graph layouts to represent attack chains, where nodes represent individual sessions and edges represent temporal or behavioral relationships between sessions. With interactive filtering, analysts can focus on certain times, types of threats, or ways of attacking. Timeline views that show how attack campaigns have changed over time and heatmap displays that show times when attacks are most active are examples of temporal visualization features.

Advanced analytics capabilities include statistical summaries of attack characteristics, comparison tools for identifying similarities between different attack campaigns, and predictive modeling features that help analysts anticipate future attack activities. The interface integrates with external threat intelligence platforms, automatically enriching displayed information with context from public and private threat feeds. Collaboration features enable multiple analysts to simultaneously investigate threats, with shared annotations and discussion threads attached to specific attack chains or sessions.

\subsection{Automated Bot Versioning and Campaign Tracking}
\noindent
Based on empirical analysis of MADCAT and T-Pot traffic, it was observed that most externally exposed sessions are generated by automated agents ("bots"), whereas sustained human interaction is comparatively rare. Accordingly, in the \emph{future architecture} the capability set is envisioned to extend beyond anomaly detection and selective escalation: recurring \emph{bot attack chains} are to be extracted automatically, clustered into families, and versioned. The intended outcome is a continuously updated, machine-generatable catalog of bot families and versions that can be referenced, detected, and shared.

\needspace{4\baselineskip}
\subsubsection{From Sessions to Canonical Attack Chains}
In the envisioned architecture, high-interaction pods are intended to capture full multi-step behaviors. Each session would be converted into a canonical attack chain by:
\begin{itemize}
  \item \textbf{Normalization}: tokenize and normalize commands (lowercasing, variable masking, path canonicalization), de-duplicate transient markers (timestamps, PIDs), and abstract parameters (e.g., IPs/URLs to classes).
  \item \textbf{Sequencing}: order events by causality and time; merge retries; compress short loops; preserve protocol boundaries (SSH/HTTP/Telnet, etc.).
  \item \textbf{Graphization}: represent the chain as a labeled multigraph (nodes: actions/resources; edges: temporal/causal), which is robust to benign reordering.
\end{itemize}

\needspace{4\baselineskip}
\subsubsection{Similarity, Clustering, and Family Detection}
An online clustering pipeline is proposed to group chains into \emph{bot families}:
\begin{itemize}
  \item \textbf{Sequence/Graph similarity}: dynamic-time-warping over action sequences and motif/subgraph similarity over the graph form.
  \item \textbf{Feature embedding}: bag-of-actions + n-grams + graph kernels, projected into a vector space for approximate nearest-neighbor search.
  \item \textbf{Incremental clustering}: streaming DBSCAN/HDBSCAN for density-based grouping with outlier discovery, yielding stable families with drift tolerance.
\end{itemize}
Clusters would be periodically stabilized into \emph{families}. A family would be promoted when support (number of chains, temporal span, multi-sensor presence) exceeds thresholds and when core motifs remain stable under canonicalization.

\needspace{4\baselineskip}
\subsubsection{Cross-Session Stitching Across IPs}
To counter typical evasion tactics in which an operator rotates source IP addresses or alternates short bursts of activity across sessions, cross-session stitching is envisioned. Sessions from different source IPs and even different sensors would be linked into a single, coherent attack chain when probabilistic evidence supports a common origin. Evidence sources would include: (i) high-similarity command subsequences and rare action motifs, (ii) shared infrastructure indicators (domains, URLs, file hashes, SSH fingerprints), (iii) consistent environment-discovery behaviors (e.g., identical probing order, idiosyncratic typos/flags), and (iv) temporal proximity within configurable windows. A session-linking model (e.g., a graph-based linkage with learned edge weights or a sequence labeling approach) would assign link probabilities; chains exceeding a confidence threshold would be aggregated into one multi-session attack chain. In this way, bursty actions issued from multiple IPs are expected to be reconstructed into a single campaign entity, enabling accurate family assignment and versioning.

\needspace{4\baselineskip}
\subsubsection{Versioning Model}
Within a family, semantically meaningful versions are to be assigned based on chain changes. A three-tier scheme inspired by semantic versioning is proposed:
\begin{itemize}
  \item \textbf{MAJOR}: structural change of the attack graph (new stage added/removed/reordered, new capability such as lateral movement, binary delivery vector change). Detected via motif divergence or graph edit distance above a major threshold.
  \item \textbf{MINOR}: stage preserved, but different tools/URLs/encodings or altered parameters/decoders; command templates change while intent remains. Detected via template delta beyond a minor threshold.
  \item \textbf{PATCH}: cosmetic or operational tweaks (timing, sleep, benign obfuscation) that do not affect capabilities; detected via low-impact deltas.
\end{itemize}
To ensure global uniqueness and operational clarity, it is proposed that each bot be named as:
\begin{center}
\texttt{\seqsplit{ADLAH.BOT.<family>.<proto>.<vector>.<platform>:<MAJOR>.<MINOR>.<PATCH>}}
\end{center}
\noindent where \texttt{<family>} is a short, mnemonic identifier promoted from the dominant cluster label (e.g., \texttt{mirailike}); \texttt{<proto>} encodes the primary ingress protocol (e.g., \texttt{SSH}, \texttt{TELNET}, \texttt{HTTP}); \texttt{<vector>} denotes the first weaponization step (e.g., \texttt{wget-curl}, \texttt{bruteforce}, \texttt{redis-rce}); and \texttt{<platform>} captures the observed target OS/arch when inferable (e.g., \texttt{linux-arm}).\\
\noindent The version tuple follows the rules above.\\
\noindent Examples: \texttt{\seqsplit{ADLAH.BOT.mirailike.TELNET.bruteforce.linux:2.1.0}}.

\needspace{4\baselineskip}
\subsubsection{Change Detection and Version Bumping}
New chains would be compared against a family's canonical representatives. If graph edit distance or template deltas were to cross MINOR/MAJOR thresholds, a new version would be proposed and audited automatically in the analyst UI. \emph{Patch} increments may be auto-accepted to reduce analyst load. Thresholds are intended to be learned from historical variance per family to avoid over-fragmentation.

\needspace{4\baselineskip}
\subsubsection{Detection and Export}
Once a family/version was established, the following artifacts would be automatically materialized:
\begin{itemize}
  \item \textbf{Online signatures}: light-weight behavioral rules for the Sensor Node and first-flight filters (e.g., action subsequences, header fingerprints) to triage and route.
  \item \textbf{IDS/YARA/Rule exports}: export to Snort/Suricata and YARA-L, linked to family/version identifiers.
  \item \textbf{TI artifacts}: compact STIX objects with family/version metadata and canonical chains for sharing.
\end{itemize}

\needspace{4\baselineskip}
\subsubsection{RL Integration}
Families and their novelty scores would feed back into the RL loop: (i) escalation would be promoted when a chain is closest to unknown families; (ii) pods with richer telemetry would be prioritized for uncertain families; (iii) confirmation of MINOR/MAJOR changes would be rewarded. This coupling is intended to steer resources toward emerging variants while keeping familiar bots inexpensive to process.

\needspace{4\baselineskip}
\subsubsection{Why Versioning Matters}
In a bot-dominated exposure model, versioning is expected to provide: (i) stable referents for operators and partners; (ii) immediate rules for early triage; (iii) quantitative measures of ecosystem drift; and (iv) a foundation for longitudinal intelligence (campaign continuity, infrastructure reuse, kit lineage). The capability described in this section forms part of the planned, future architecture rather than the present prototype.

\needspace{5\baselineskip}
\subsection{Overview and Design Principles}
\noindent
The adaptive honeynet architecture is built upon several key design principles that ensure effective threat detection, resource efficiency, and operational flexibility. The system employs a modular design that separates concerns across distinct components, enabling independent development, testing, and deployment of each subsystem while maintaining seamless integration through well-defined interfaces. This modular approach facilitates maintenance, troubleshooting, and future enhancements while reducing the complexity of individual components.

Scalability is achieved through containerized deployment and cloud-native architecture, enabling the system to scale resources based on threat volume and computational demands dynamically. The containerized approach ensures consistent deployment across various environments, while allowing for rapid scaling and resource optimization. Security is maintained through network isolation, secure communication channels, and comprehensive logging that provides visibility into all system activities while protecting against potential compromise of the honeynet infrastructure itself.

The architecture prioritizes real-time processing capabilities, enabling immediate response to detected threats through rapid pod deployment and traffic redirection. This real-time capability is essential for maintaining the illusion of vulnerable systems that adversaries expect to encounter, ensuring that honeypots are available when needed to capture valuable threat intelligence. The system also emphasizes adaptability through machine learning integration, allowing it to evolve its threat detection and response capabilities based on observed attack patterns and system performance.~\footnote{This is often referred to as online or continuous learning, where the model updates itself incrementally as new data arrives, in contrast to static models trained offline.}

The architecture is fundamentally designed to scale from its current single-sensor prototype to a large, geographically distributed honeynet. This scalability is rooted in the clear separation of the lightweight \textbf{Sensor Nodes} from the central \textbf{Hive}. Multiple sensor nodes can be deployed across diverse networks and regions, each acting as an independent data collector and enforcement point. These sensors report threat telemetry to a central Hive, which aggregates intelligence and centralizes the computationally intensive AI/ML workloads. The use of Kubernetes for high-interaction honeypot deployment is another key enabler of scale; as the number of required engagements grows, the cluster can be expanded by adding more worker nodes. This distributed model not only broadens threat visibility but also creates a resilient infrastructure capable of supporting large-scale intelligence operations without creating performance bottlenecks at the edge.
\subsection{Sensor Node}
The Sensor Node runs MADCAT, a low-interaction honeypot developed by the BSI that records all connection attempts without committing to particular services~\cite{bsi2025madcat-3e8}. While MADCAT provides a crucial baseline for threat observation, this architecture elevates it from a passive sensor to an intelligent tripwire for a dynamic, adaptive system. By integrating a sophisticated AI stack, this work directly supports the BSI's strategic goals: to accelerate the detection of new attacks and to gain more profound, more actionable insights into application-layer threats. This enhanced understanding is vital for the BSI's mission to produce timely and effective public warnings and advisories. MADCAT opens all ports and accepts Transmission Control Protocol (TCP)~\cite{ed2022transmission-63c}, User Datagram Protocol (UDP)~\cite{postel1980user-caf}, and raw traffic, providing only the technically necessary responses to establish data connections with potential adversaries. This low-interaction approach enables the capture of attack vectors while minimizing the risk of detection. The Sensor Node is deployed as a container on a Virtual Machine (VM) with two network interfaces: one for adversary-facing traffic and one for internal Hive communication. The MADCAT configuration has been modified to log first-packet metadata (headers, payload, timestamps) to `/var/log/madcat`, which is then forwarded via Logstash over Transport Layer Security (TLS) to the Hive node.

\subsection{Hive Node}
The Hive node hosts an ELK (Elasticsearch, Logstash, Kibana) stack for centralized data processing~\cite{elasticnoyearelastic-42b}. Logstash receives structured MADCAT logs via TCP~\cite{ed2022transmission-63c}, applies parsing filters, and indexes them into Elasticsearch. Kibana is secured behind a reverse proxy and used for interactive analysis. The RL decision agent also runs on the Hive and continuously polls Elasticsearch for recent sensor events to drive deployment decisions in real time.

\subsection{Reinforcement Learning Agent}
\label{sec:rl-orchestration}
The RL agent runs as a containerized Python service on the Hive node. It consumes recent events via an Elasticsearch polling source and builds short sequences per source IP. The agent is a TensorFlow-based Deep Q-Network with an LSTM encoder and a dueling head (separate value and advantage streams) optimized with Adam (gradient clipping) and Huber loss~\cite{abadi2016tensorflow-680}. It uses experience replay, a target network, and an \(\epsilon\)-greedy policy. The discrete action space is \{\texttt{wait}, \texttt{deploy}\}. Decisions are executed directly via the Kubernetes API client, which creates a labeled honeypot deployment; the sensor side is configured for forwarding by a controlled bootstrap mechanism. For the prototype, a constrained, non-interactive SSH channel is used over key-based auth with host verification; a production deployment should replace this with an agent-based approach (e.g., mTLS control plane or SSM-like mechanism) to reduce attack surface. When cluster utilization is low, the event loop may promote a \texttt{wait} decision to \texttt{deploy} (heuristic: promote if max(CPU, memory) < 0.6; configurable) to avoid idle capacity. Rewards are assigned asynchronously: successful honeypot hits yield positive rewards; expired deployments or deploys skipped due to resource pressure incur small negative rewards. Advanced anomaly detection is part of the roadmap, but not active in the prototype.

\subsection{Cluster Node and Pod Deployment}
The cluster node runs a lightweight `k3s` Kubernetes installation~\cite{noauthornoyearlightweight-bd4} for the prototype; in a managed production setting, GKE or a full Kubernetes distribution would be preferable. It receives deployment instructions from the Hive's RL agent and instantiates honeypot pods in isolated namespaces. Pods are labeled with the attacker's source IP to support later lookup of their assigned pod IP. Traffic steering is not performed in the cluster; instead, the event loop configures sensor-side forwarding to the pod IP for the relevant ports
and links it back to the source IP of the session. Pods are preloaded and then, after the deploy decision, renamed, and the traffic gets forwarded to them. Utilizing pre-loaded base images is a deliberate countermeasure against detection via deployment latency, as it minimizes the provisioning delay that could be fingerprinted by an adversary.

\subsection{Traffic Forwarding and Redirection}
A core challenge is transparently redirecting adversary traffic from the low-interaction Sensor Node to a high-interaction honeypot on the cluster node without alerting the adversary. The system must maintain session integrity while redirecting traffic between the distributed components. The prototype implementation, detailed in Section~\ref{sec:prototype}, uses dynamic \texttt{iptables} rules to achieve this, ensuring that the adversary remains unaware of the infrastructure transition.

\subsection{Integration with Traditional Intrusion Detection Systems}
The architecture is designed for extensibility and can be integrated with traditional signature-based Intrusion Detection Systems (IDS) such as Snort~\cite{cisconoyearsnort-193}. This integration creates a synergistic defense mechanism that combines the strengths of both behavioral analysis and signature matching. Snort can be deployed on the Sensor Node, where it would monitor inbound traffic in parallel with MADCAT.

Alerts generated by Snort can serve as a valuable, additional input for the RL agent. The state vector for the DQN could be augmented with features derived from Snort alerts, such as signature priority or threat classification. This would provide the agent with a richer context, allowing it to make more informed decisions about whether to deploy a high-interaction honeypot. For instance, a high-priority Snort alert can serve as a strong signal to escalate a session for further analysis.

Furthermore, Snort logs can be forwarded to the ELK stack on the Hive node. This enables analysts to correlate signature-based alerts with the behavioral data captured by the honeypots, all within a single Kibana dashboard. This unified view is critical for comprehensive threat analysis. In a more advanced implementation, a feedback loop could be established where the AI Analytics Pipeline analyzes novel attack patterns from the high-interaction pods and automatically generates new Snort rules. This would enable the adaptive honeynet to harden the traditional IDS against zero-day threats dynamically.

\section{Methodology}
\label{sec:methodology}
\noindent
This section details the research methodology, including the experimental design, data collection procedures, and data analysis techniques.

\subsection{Data Collection and Environment}
The primary dataset is sourced from the MADCAT Sensor Node's JSON-based logging output, aggregated in an Elasticsearch index. Each document in the index represents a network event. The prototype was deployed in a live environment connected to the public internet and exposed to genuine, unsolicited traffic.

In parallel, the possibility of obtaining T-Pot data from Telekom Security for research use was explored in coordination with BSI. Due to the high volume of external requests and associated review processes, access was not feasible within the time frame of this thesis. To partially compensate, an independent T-Pot instance was provisioned, and exploratory data were collected. These data are not used in the present evaluation but are intended to inform future versions of the architecture, especially for application-layer anomaly detection and model pretraining.

\subsection{Data Analysis and Feature Engineering}
The foundation of any machine learning-driven security system is the data it consumes. The raw data generated by the MADCAT Sensor Node, while comprehensive, is not immediately suitable for direct input into the reinforcement learning agent. This section details the critical process of data analysis, cleaning, and feature engineering required to transform raw log entries into a meaningful state representation for the RL agent.

\subsubsection{Raw Data Characteristics}
An initial analysis of the dataset reveals several essential characteristics. The data exhibits high cardinality and sparsity, containing more than one hundred unique fields, many of which are protocol-specific and therefore present in only a subset of records. The features span mixed data types, including numerical, categorical, boolean, and complex string-based values, with many numerical fields requiring type conversion. Significant redundancy is also present, for example, in duplicated keyword fields, and missing values occur frequently, necessitating systematic handling.

\subsubsection{Data Preprocessing and Cleaning}
Before feature engineering, the raw data passes through a rigorous cleaning and preprocessing pipeline. Numerical features are converted to their correct data types, while missing values are handled by imputing zeros for numerical fields and introducing a dedicated 'missing' category for categorical variables. Categorical features with low cardinality are one-hot encoded to ensure compatibility with downstream models. Temporal information is extracted from timestamps in the form of cyclical features, enabling the learning of temporal patterns such as diurnal or weekly attack behaviors.

\section{Prototype Implementation and Demonstration}
\label{sec:prototype}
The prototype implements the core components of the RL-driven orchestration loop: the RL agent itself, the integration with the MADCAT sensor, and the dynamic pod deployment and redirection mechanism. This section details the practical implementation of these components, demonstrating the feasibility of the proposed approach and providing a solid foundation for the future enhancements outlined in the architectural vision.

\subsection{Implementation Details}
\label{sec:implementation_details}
\noindent
The prototype was built using a combination of open-source tools and custom scripts, deployed on Google Cloud Platform. The complete source code, deployment scripts, and documentation are available in the project's GitHub repository at \url{https://github.com/JohannesLks/adlah.git}.

In addition, prebuilt Docker images are published on Docker Hub to facilitate quick setup and reproducibility. Table~\ref{tab:docker_images} lists the main images for the sensor and hive components.

\begin{table}[h]
\caption{Published Docker Images on Docker Hub\label{tab:docker_images}}
\centering
\renewcommand{\arraystretch}{1.15}
\begin{tabularx}{\linewidth}{l X}
\hline
\textbf{Component} & \textbf{Docker Hub Repository} \\
\hline
Sensor & \href{https://hub.docker.com/r/lukasjohannes/adlah-sensor-madcat}{adlah-sensor-madcat} \\
Sensor & \href{https://hub.docker.com/r/lukasjohannes/adlah-sensor-filebeat}{adlah-sensor-filebeat} \\
Hive & \href{https://hub.docker.com/r/lukasjohannes/adlah-hive-elasticsearch}{adlah-hive-elasticsearch} \\
Hive & \href{https://hub.docker.com/r/lukasjohannes/adlah-hive-kibana}{adlah-hive-kibana} \\
Hive & \href{https://hub.docker.com/r/lukasjohannes/adlah-hive-rl-agent}{adlah-hive-rl-agent} \\
Hive & \href{https://hub.docker.com/r/lukasjohannes/adlah-hive-logstash}{adlah-hive-logstash} \\
Hive & \href{https://hub.docker.com/r/lukasjohannes/adlah-hive-nginx}{adlah-hive-nginx} \\
\hline
\end{tabularx}
\end{table}

\begin{figure}[h]
\centering
\includegraphics[width=0.9\columnwidth]{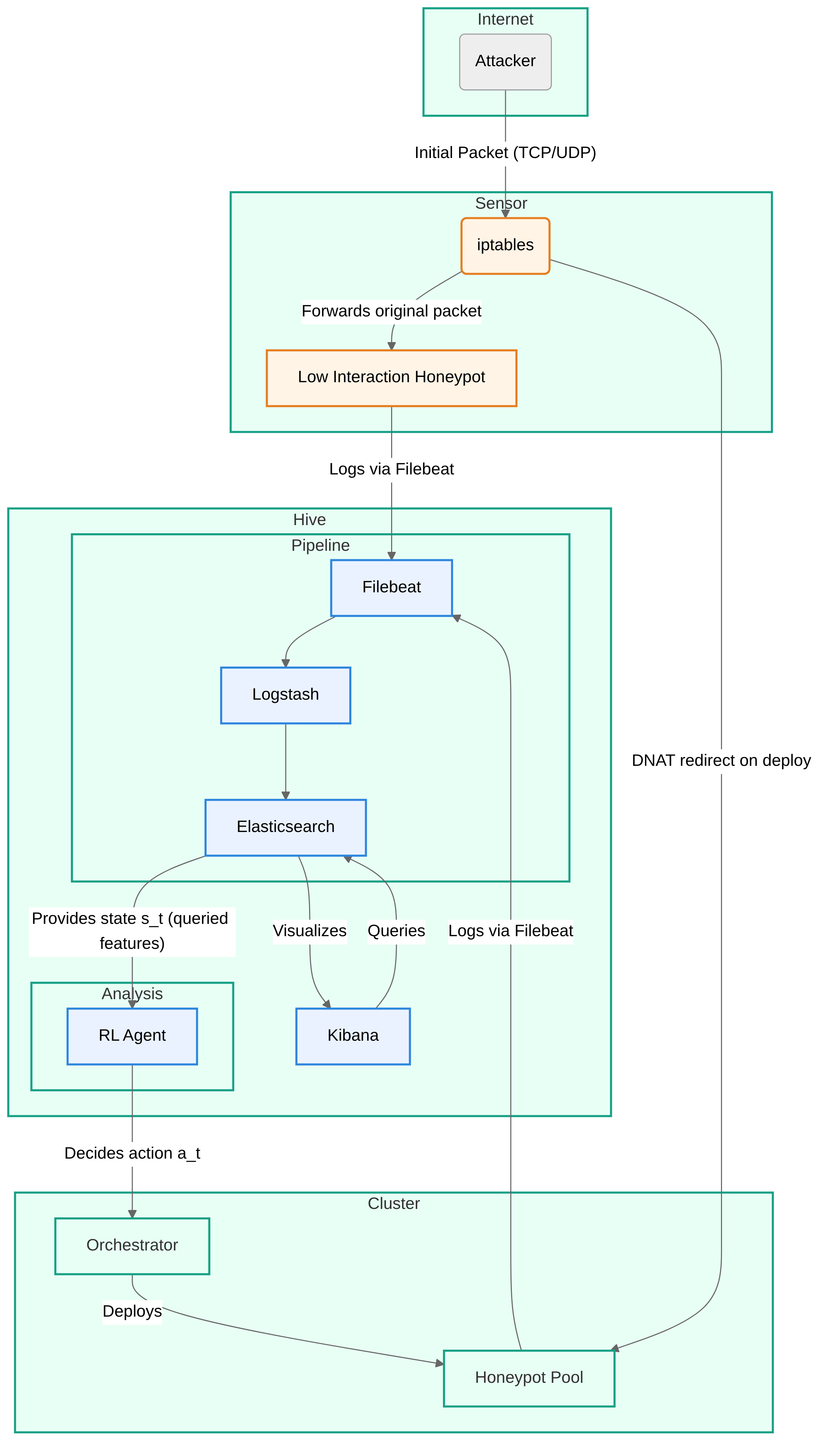}
\caption{Visual overview of the prototype's RL-Agent architecture.}
\label{fig:prototype-rl-agent-architecture}
\end{figure}

The Sensor Node is implemented using a modified MADCAT configuration to enhance logging for the RL agent. The Hive node runs an ELK stack (Elasticsearch, Logstash, Kibana) for centralized data processing and hosts the RL agent. The cluster node uses `k3s` for lightweight Kubernetes orchestration. When the agent triggers a deployment, it instantiates an SSH honeypot (for example, Cowrie~\cite{teamnoyearcowrie-975}). A visual overview of the RL-Agent architecture and data flow is provided in Figure~\ref{fig:prototype-rl-agent-architecture}, which illustrates the integration between the RL decision-making loop, container orchestration, and log aggregation.

The system state implemented in the prototype demonstrates an operational end-to-end loop for first-flight detection and decision-making. However, the attacker forwarding mechanism, which was planned to use WireGuard\cite{donenfeldnoyearwireguard-ffb} for secure tunneling, could not be successfully implemented despite extensive effort. This limitation prevented the collection of high-interaction (deep emulation) session data, which is essential for validating the anomaly detection pipeline and the envisioned quality-weighted reward function. Although a DNAT-based forwarding mechanism later became available, it arrived too late in the development cycle to produce a fully running proof-of-concept with post-escalation data capture.

A custom Bash script, \texttt{reinstall.sh}, handles the complete setup for all nodes.

\subsection{Data-Driven Agent Design: A Foundational Analysis}
\noindent
To ensure the reinforcement learning agent's architecture is grounded in real-world conditions, a comprehensive preliminary analysis was conducted on a large-scale dataset provided by the German Federal Office for Information Security (BSI). This dataset, captured by their MADCAT Sensor Node network throughout November 2024, contains millions of log entries and represents a realistic snapshot of typical internet background noise and unsolicited traffic. This analysis was not merely for environmental context, but served as a foundational step in designing the agent's core logic.

The statistical properties of this data directly informed key architectural decisions. For instance, the distribution of events per source IP was heavily skewed, as shown in Table~\ref{tab:dataset_statistics}. While the average number of events per IP was high due to aggressive scanners, the median was significantly lower. The median of 4.0 interactions per IP indicates that at least half of the source addresses appear at most four times. This guided a design that supports early decisions on short sequences while still modeling longer interactions: the agent consumes up to the last \(N=10\) observations with masking of zero-padded positions.

\begin{table}[h!]
\caption{Summary of BSI MADCAT Network Traffic (Nov. 2024)}
\label{tab:dataset_statistics}
\centering
\renewcommand{\arraystretch}{1.2}
\begin{tabularx}{\linewidth}{l X}
\hline
\textbf{Metric} & \textbf{Value} \\
\hline
Total Log Events & 13,244,068 \\
Unique Source IPs & 133,429 \\
Analysis Period & November 1, 2024 - November 30, 2024 \\
\hline
\multicolumn{2}{l}{\textit{\textbf{Monthly per-IP Frequency (whole month)}}} \\
Mean per-IP Frequency (month) & 99.26 \\
Median per-IP Frequency (month) & 4.0 \\
\hline
\multicolumn{2}{l}{\textit{\textbf{Daily per-IP Frequency (by active IPs on that day)}}} \\
Day with Highest Daily Mean &  2024-11-07 (per-IP daily mean: 36.18) \\
Day with Lowest Daily Mean & 2024-11-15 (per-IP daily mean: 19.54) \\
\hline
\end{tabularx}
\end{table}

\noindent\textit{Note:} The monthly mean is computed as total events divided by the number of unique IPs across the whole month. Daily means are computed per day over the set of IPs active on that day. These aggregates are therefore not directly comparable and serve different analytical purposes.

\subsection{Reinforcement Learning Agent Design}
\label{sec:rl_agent_design}
\noindent
The reinforcement learning component orchestrates honeypot deployments from short sequences of recent events per source IP. This section reflects the current implementation: an LSTM-based dueling DQN with experience replay, a target network, and asynchronous reward assignment, integrated with Kubernetes for pod lifecycle and with the Sensor Node for per-attacker forwarding.

\subsubsection{State Space (\texorpdfstring{$\mathcal{S}$}{S})}
The state, $s_t \in \mathcal{S}$, is designed to capture temporal patterns in network traffic by representing a sequence of the last $N$ observations from a single source IP address.

\needspace{4\baselineskip}
\subsubsection{Sequence Length and Early Decisions}
The state is a fixed-length sequence of the most recent $N=10$ observations. If fewer than $N$ observations are available for a given IP, the sequence is zero-padded and masked so that the LSTM ignores padded steps. The agent is invoked from the first observation onward; padding permits decisions at $t\in\{1,\dots,N\}$ without waiting for a full sequence window. This choice reflects the empirical median length (\(\approx 4\)) and the long tail of higher activity levels.
\needspace{4\baselineskip}
\subsubsection{Observation Features}
Each observation $o_t$ is a fixed-length feature vector derived from a single parsed network event in Elasticsearch, restricted to fields that are directly usable by the RL agent in real-time without heavy external enrichment or preprocessing. The feature composition in the current prototype is:

\begin{itemize}
    \item \textbf{Core numeric features (6):} \code{FLOW.duration} (s), \code{FLOW.bytes\_toserver}, \code{FLOW.bytes\_toclient}, \code{IP.ttl}, \code{TCP.dest\_port}, \code{TCP.src\_port}. These fields capture timing, data volume, and port targeting, primary indicators for malicious session behavior.
    
    \item \textbf{ASN numeric feature (1):} \code{src\_ip\_geoasn.as.number}. The Autonomous System Number is directly used for coarse-grained reputation scoring without requiring organization-name resolution.
    
    \item \textbf{Categorical features with hashed one-hot encoding (140 dims total):}
    \begin{itemize}
        \item \code{src\_cc} (32 dims) ISO country code of source IP, when available without enrichment delay.
        \item \code{tcp\_flag\_pattern} (32 dims) compact encoding of observed TCP flag combinations.
        \item \code{FLOW.state} (16 dims) protocol-independent connection state from flow tracking.
        \item \code{event\_type} (32 dims) parsed event classification (e.g., \texttt{flow}, \texttt{syn\_scan}).
        \item \code{proto} (8 dims) transport protocol category.
        \item \code{port\_category} (20 dims) fixed mapping for well-known, registered, dynamic, and specific high-value ports (\texttt{22}, \texttt{23}, \texttt{80}, \texttt{443}, \texttt{5060}, \texttt{123}).
    \end{itemize}
    High-cardinality text values (e.g., AS names) are excluded to avoid unnecessary collisions and enrichment cost. Hashed one-hot encoding using \texttt{xxhash} ensures a fixed dimensionality with stable mapping for categorical domains of bounded size.
    
    \item \textbf{Short-term activity profile (5):} Computed over a rolling 5-minute window per source IP: \code{ev\_5m} (event count), \code{ports\_5m} (unique destination ports), \code{protos\_5m} (unique protocols), \code{targets\_5m} (unique destination IPs), and \code{syn\_ratio\_5m} (ratio of SYN packets to total TCP packets).
    
    \item \textbf{Time features (2):} Sine and cosine transforms of the time-of-day from \code{@timestamp}, scaled to $[0, 2\pi]$ radians, enabling cyclic encoding of diurnal patterns.
    
    \item \textbf{Runtime feature (1):} \code{time\_since\_last\_min} capped minutes since the last event from this source IP, providing a recency signal without storing the full history.
    
    \item Payload features (8 dims): Four scalar statistics computed directly on available payloads: total length, byte entropy, non-printable character ratio, and LZ77-based compressibility. The latter is derived from the compression ratio obtained by applying the Lempel-Ziv 77 algorithm~\cite{ziv1977universal-81e}, a lossless dictionary-based scheme that replaces repeated sequences with backward references. This metric serves as a heuristic for payload structure: highly compressible payloads often contain redundant or patterned data (e.g., repeated commands or cleartext), whereas low-compressibility payloads are more likely to be encrypted or obfuscated. A matching 4-dimensional block encodes the SHA-1 hash of the payload (truncated to 32-bit integers for space efficiency). Raw payload bytes and strings are not retained; if no payload is present, the payload sub-vector is set to all zeros.
\end{itemize}

This configuration intentionally excludes enrichment-heavy features (e.g., full geolocation, OS fingerprinting, organization names) and non-deterministic metadata. The design ensures all inputs are available from parsed event fields at ingest time, keeping the observation extraction path bounded in latency and reproducible.

\needspace{4\baselineskip}
\subsubsection{Processing and Normalization}
Numeric features are normalized in an online fashion using Welford's algorithm~\cite{welford1962note-677} to incrementally compute mean and variance, with hard clipping applied to prevent extreme values from dominating the feature scale. Running statistics are persisted to disk so that scaling remains consistent across process restarts. Categorical variables are transformed into fixed-size indicator vectors via deterministic hashing with \texttt{xxhash}~\cite{cyan4973noyearxxhash-fdc}, enabling constant-time transformation while bounding dimensional growth. The \code{time_since_last_min} feature is stored in its own dedicated scalar slot rather than as part of any hashed indicator vector, ensuring that recency information remains numerically interpretable and does not collide with categorical encodings.

See Appendix~\ref{app:appendix_featured_analysis} for an analysis of raw MADCAT fields. The exact model input features used by the RL agent are listed in Section~\ref{sec:rl_agent_design} (Observation Features).

\subsubsection{Action Space (\texorpdfstring{$\mathcal{A}$}{A})}
The agent's action space is discrete and consists of two possible actions. The action $a_t \in \mathcal{A}$ is chosen at each time step $t$:
\begin{enumerate}
    \item \texttt{wait}\,: The agent takes no action and continues to observe events from the source IP. This is the default, resource-conserving action.
    \item \texttt{deploy}\,: The agent initiates the deployment of a high-interaction honeypot tailored to the perceived threat and redirects the source IP's traffic to it.
\end{enumerate}

\subsubsection{Reward Function (\texorpdfstring{$\mathcal{R}$}{R})}
The reward signal $R(s_t, a_t)$ is defined to favor deployments that actually yield intelligence. Rewards are sparse and delayed, reflecting that utility is only observable after a honeypot's observation window.

\needspace{4\baselineskip}
\subsubsection{Current Design (Prototype)}
If the agent chooses deploy, the reward is proportional to the amount of log data collected from the targeted source after deployment. Let $L$ denote the total number of log events attributed to that source during the honeypot's lifespan, and $\bar{L}$ a normalization constant (e.g., a rolling median per protocol/port or per time-of-day bucket). The deployment reward is:
\begin{equation}
R_{deploy} = \alpha \cdot \min\left(\frac{L}{\bar{L}}, L_{max}\right)
\end{equation}
with $\alpha > 0$ a scaling factor and $L_{max}$ a cap to limit outliers. If no logs are collected ($L = 0$), a mild penalty is assigned to discourage wasteful deployments:
\begin{equation}
R_{no\_logs} = -\delta, \quad \delta \ll 1
\end{equation}

To optimize resource utilization and discourage idle deployments, each high-interaction pod in the cluster is configured with a 20-minute inactivity timeout. 
If no interactions from the associated adversary are observed within this window, the pod is automatically terminated. 
Such premature termination triggers a penalty in the RL agent's reward function, reflecting the wasted resources and missed intelligence opportunities. 
This mechanism incentivizes the agent to escalate only when there is a high likelihood of sustained interaction, aligning operational efficiency with intelligence-gathering objectives.

\needspace{4\baselineskip}
\subsubsection{Planned Extension (Quality-Aware Reward)}
The reward will be augmented with a novelty/anomaly term that reflects the quality of the collected data. Let $A$ denote an anomaly score aggregated over the new logs (e.g., mean reconstruction error from an autoencoder, or an alternative detector's score). The future reward will take the form:
\begin{equation}
R_{deploy}^{future} = \alpha \cdot \min\left(\frac{L}{\bar{L}}, L_{max}\right) + \beta \cdot \text{clip}(\text{Agg}(A), 0, A_{max})
\end{equation}
where $\beta > 0$ weights anomaly-driven value, $\text{Agg}(\cdot)$ is a robust aggregator (e.g., trimmed mean or percentile), and $A_{max}$ caps the anomaly contribution. This preserves a conservative shaping philosophy: intelligence quantity remains the backbone, while quality signals (novelty/anomaly) enhance learning without enabling reward gaming. The utilization heuristic continues to act only as a stabilizer outside the reward to avoid starvation during low-interaction phases.

\subsubsection{Learning Algorithm}
The agent employs a Deep Q-Network (DQN) with an LSTM layer to process the sequential state information.

\needspace{4\baselineskip}
\subsubsection{Neural Network Architecture}
The network architecture is comprised of the following layers (resource-awareness noted where applicable):
\begin{enumerate}
    \item \textbf{Input Layer:} Takes the state tensor of shape $(10, D)$, where $D$ is the feature dimension.
    \item \textbf{Masking Layer:} Handles variable-length sequences by ignoring padded time steps, ensuring they do not affect the LSTM's state.
    \item \textbf{LSTM Layer:} A single LSTM layer with 64 units processes the sequence, capturing temporal dependencies in the data.
    \item \textbf{Batch Normalization:} Stabilizes learning by normalizing the activations from the LSTM layer.
    \item \textbf{Dense Layer:} A fully-connected layer with 64 units and a ReLU activation function. Runtime features (cluster utilization, active pods, capacity) are concatenated post-LSTM to allow explicit conditioning of Q-values on resource pressure.
    \item \textbf{Dropout Layer:} A dropout rate of 0.2 is applied for regularization to prevent overfitting.
\item \textbf{Dueling Head:} Two parallel streams compute (i) the state value and (ii) the action advantages. They are combined into Q-values via \(Q(s,a)=V(s)+A(s,a)-\tfrac{1}{|\mathcal{A}|}\sum_{a'}A(s,a')\).~\cite{wang2016dueling-92c} The output comprises two Q-values, one per action (\texttt{wait}, \texttt{deploy}).
\end{enumerate}

\needspace{4\baselineskip}
\subsubsection{Q-Learning Update Rule}
To improve the stability of value estimation, the agent combines two mechanisms from prior work: 
(i) \emph{experience replay}~\cite{mnih2015human-level-aca} to decorrelate updates by sampling from a replay buffer, and 
(ii) a \emph{target network}~\cite{mnih2015human-level-aca} to reduce non-stationarity in the bootstrap target.
The target network parameters $\theta_t'$ are held fixed for a number of training steps and then updated from the online network parameters $\theta_t$.

The action-value target follows the \emph{Double DQN} formulation~\cite{hasselt2016deep-2f7}, in which the next action is selected using the online network and evaluated using the target network. 
For a transition $(s_t, a_t, R_{t+1}, s_{t+1}, \textit{done}_{t+1})$, the target is:
\begin{equation}
\begin{split}
    y_t &= R_{t+1} + (1 - \textit{done}_{t+1}) \, \gamma \\
        &\quad \cdot Q\!\left(s_{t+1}, \arg\max_{a} Q(s_{t+1}, a; \theta_t);\, \theta_t' \right),
\end{split}
\end{equation}
where $\gamma \in [0,1]$ is the discount factor. 
The multiplicative $(1 - \textit{done}_{t+1})$ term is not explicitly shown in~\cite{hasselt2016deep-2f7} but is standard in implementations to zero out the bootstrap term when the next state is terminal.

Following~\cite{mnih2015human-level-aca}, the network parameters $\theta_t$ are optimized by minimizing the \emph{Huber loss} between the predicted $Q(s_t,a_t;\theta_t)$ and the target $y_t$, which is less sensitive to outliers than the mean-squared error. 
This formulation mitigates the overestimation bias of standard Q-learning~\cite{watkins1992q-learning-821} while retaining the sample efficiency of off-policy learning.

\needspace{4\baselineskip}
\subsubsection{Hyperparameters}
The key hyperparameters for the learning process are:
\begin{itemize}
    \item Learning Rate ($\alpha$): 0.001
    \item Discount Factor ($\gamma$): 0.95
    \item Initial Epsilon ($\epsilon$): 1.0 $\rightarrow$ linear decay to $\epsilon_{min}$ (for robust early exploration)
    \item Minimum Epsilon ($\epsilon_{min}$): 0.01
    \item Epsilon Decay: 0.995
    \item Replay Memory Size: 10,000
    \item Batch Size: 32
    \item Target Network Update Frequency: 1000 training steps (hard update)
\end{itemize}

These hyperparameters represent initial starting values for training and are not guaranteed to be optimal for the target environment. 
Determining their effectiveness requires empirical evaluation after system deployment, when sufficient interaction data are available for the RL agent to operate in a fully online setting. 
As part of future work, it is planned to apply automated hyperparameter tuning techniques, such as Bayesian optimization~\cite{snoek2012practical-888}, to identify the configuration that maximizes long-term performance.

More broadly, all design elements described in this section, including feature selection, hashing strategies, normalization schemes, and reward shaping, represent the current prototype implementation. 
They are expected to undergo significant refinement as empirical evidence is gathered, and the final deployed system will likely differ substantially from the present description (see Section~\ref{sec:current_limitations}).

\subsection{Deployment and Infrastructure (Prototype)}
\noindent
This section documents the current prototype environment and clarifies the handling of cluster utilization. Maintaining a healthy utilization band improves learning efficiency and intelligence yield: if utilization remains too low due to overly conservative deployment (as observed in early tests with sparse sensor interactions), the agent receives too few meaningful rewards; if too high, contention harms fidelity.

\subsubsection{Cloud Infrastructure}
The cloud infrastructure is deployed on Google Cloud Platform (GCP), taking advantage of its scalable and flexible computing resources. 
Compute Engine virtual machines (VMs) host the sensor, Hive, and cluster nodes, with each instance provisioned according to the specific performance requirements of its role. 
This right-sizing approach ensures optimal operational performance while controlling costs. 
Container images for the deployment are stored and versioned in GCP's managed Container Registry, providing a secure and centralized repository for maintaining the integrity and traceability of all containerized components.

\subsubsection{Network Configuration}
The network configuration ensures secure and efficient communication between the different components of the honeynet:
\begin{itemize}
    \item \textbf{VPC and Subnets:} A Virtual Private Cloud (VPC) with subnets for isolating different components of the honeynet.
    \item \textbf{Firewall Rules:} Configured to allow only necessary traffic to and from the honeynet components, following the principle of least privilege.
    \item \textbf{Routing:} Custom routes to manage the traffic flow between the sensor, Hive, and cluster nodes.
\end{itemize}

\subsection{Configuration Management}
\noindent
The prototype uses infrastructure-as-code for reproducible deployment.

\subsubsection{Terraform Configuration}
Infrastructure is managed using Terraform~\cite{hashicorpnoyearterraform-5bd}, providing a declarative approach to infrastructure provisioning that ensures consistency and reproducibility across different deployment environments. Resource definitions specify VM instances, networks, and storage components with precise configurations that can be version-controlled and audited, enabling teams to track changes and understand the impact of infrastructure modifications. This approach eliminates configuration drift, ensuring that all deployments utilize identical infrastructure configurations.

\subsubsection{Docker Compose}
Local development utilizes Docker Compose to provide a consistent and reproducible development environment that mirrors the production deployment architecture. Service definitions specify container configurations for all system components, ensuring that developers can work with the identical software versions and configurations that will be deployed in production. This consistency eliminates the "works on my machine" problem and enables developers to test their changes in an environment that closely resembles the actual deployment.

Network configuration establishes inter-service communication patterns that replicate the production network topology, allowing developers to test component interactions and integration points without requiring the whole cloud infrastructure. This local networking setup enables efficient development and debugging while maintaining the security and isolation requirements of the honeynet system. Volume management provides persistent data storage for development and testing, ensuring that valuable test data and configuration information are preserved across development sessions and enabling realistic testing scenarios that require historical data.

\subsection{Traffic Forwarding Implementation}
\noindent
Transparent redirection from the Sensor Node to a deployed honeypot is orchestrated by the event loop after a \texttt{deploy} decision. The Kubernetes manager creates a pod labeled with the attacker's source IP and exposes the service ports. The event loop then obtains the pod IP and calls a sensor-side setup script over SSH using a non-interactive command. That script configures connection tracking and port forwarding (e.g., via \texttt{iptables}/DNAT) for the specific source IP and target pod IP for a configured TTL. This design keeps latency low and confines redirection to the attacker being escalated. Cleanup of expired rules and pods is performed periodically by the event loop and the honeypot manager.

\subsection{Limitations and Future Enhancements}
\noindent
The prototype has several limitations that will be addressed in future versions.

\subsubsection{Current Limitations}
~\label{sec:current_limitations}
Known limitations include scale constraints that limit the maximum number of concurrent sessions and pod deployments that can be handled simultaneously. These limitations are primarily determined by the computational resources available and the efficiency of the current implementation, which may restrict the system's ability to handle extremely high-volume attack scenarios or coordinate large numbers of honeypot instances. Feature limitations encompass the restricted range of honeypot types and detection capabilities currently supported by the prototype, which may not cover all potential attack vectors or provide the depth of analysis required for highly sophisticated threats.

Another fundamental limitation relates to the nature of the available data. 
Throughout the thesis work, the reinforcement learning agent was deployed and operated entirely in a live environment, interacting with real adversary traffic rather than static offline captures. 
This approach is essential, as the agent's reward structure depends on genuine state transitions and information gain arising from its own deployment actions. 
However, due to technical issues in the attacker forwarding mechanism, no high-interaction (deep emulation) sessions were successfully recorded. 
Since such post-escalation data are critical for both the anomaly detection pipeline and the envisioned quality-based reward function, their absence limited the scope of the evaluation to the first-flight decision-making stage. 
While a potential data exchange with Telekom Security's T-Pot platform was explored together with BSI, the required organizational approvals could not be completed within the thesis timeline. 
As a mitigation, a self-operated T-Pot instance was deployed for exploratory collection; however, these data are reserved for future work and are not included in the present evaluation.

Performance limitations arise from real-time processing constraints that affect the system's ability to maintain optimal response times under heavy load conditions. These constraints are influenced by the computational complexity of the machine learning models and the overhead associated with container orchestration and network redirection. Security limitations reflect the basic security measures and hardening currently implemented, which may not provide the level of protection required for deployment in highly sensitive environments or against sophisticated adversaries who may attempt to compromise the honeynet infrastructure itself.

\subsubsection{Scaling \& Security}
Future enhancements include scalability improvements that will implement horizontal scaling and load balancing capabilities to handle increased traffic volumes and support larger deployments across multiple geographic regions. These improvements will enable the system to distribute computational load across various nodes and automatically scale resources based on demand, ensuring consistent performance even under extreme attack scenarios. Advanced machine learning models will be integrated (e.g., transformer-based models) for more sophisticated sequence understanding and threat prediction.

Enhanced security features will incorporate advanced threat detection and response mechanisms, including more sophisticated anomaly detection algorithms, behavioral analysis capabilities, and automated response systems that can adapt to evolving attack techniques. Integration capabilities will be expanded to include seamless connectivity with SIEM systems and threat intelligence platforms, enabling the adaptive honeynet to contribute to broader security ecosystems and share threat intelligence with other security tools and platforms.

\subsubsection{Clustering \& Anomaly Detection}
At the current stage of development, no definitive solution for the anomaly detection component has been established. 
Preliminary investigations indicate that a single analytical technique is unlikely to perform optimally across all deployment scenarios, particularly given the heterogeneity of attack behaviors and the evolving nature of adversary tactics. 
Instead, it is expected that a \emph{portfolio} of complementary methods will need to be evaluated and potentially combined into a hybrid detection pipeline. 
Such a pipeline could, for example, employ lightweight classifiers,such as decision trees or Random Forest ensembles,for rapid, low-latency triage of incoming sessions, followed by more computationally intensive deep reinforcement learning (DQN) or recurrent neural network (RNN) models for fine-grained resource allocation and escalation decisions. 
Other promising avenues include clustering algorithms (e.g., DBSCAN, HDBSCAN, or k-means variants) for unsupervised grouping of session behaviors, autoencoder-based novelty detection for identifying rare or unseen patterns, and graph-based learning approaches for correlating activity across multiple sessions and attack chains.

Given the scope of these possibilities, the evaluation process will necessarily proceed in multiple phases, beginning with experiments on offline datasets to reduce operational risk and allow rapid iteration of models and features. 
To facilitate this process and build domain-specific knowledge of the data, a collaborative research initiative between the German Federal Office for Information Security (BSI) and the University of Applied Sciences Kiel is planned. 
Within this framework, student-led projects will systematically investigate and benchmark a range of deep learning and machine learning techniques, starting with exploratory clustering to uncover latent structures in the data, progressing to supervised classification methods for labeling and triaging known attack patterns, and culminating in anomaly detection methods tailored for real-time deployment. 
The insights gained from these projects will directly inform the design of a robust, hybrid anomaly detection strategy capable of adapting to the diverse operational contexts in which the system may be deployed.

\section{Evaluation (Planned)}
\label{sec:evaluation}
\noindent
Given that the reinforcement learning agent's reward structure depends on live interaction, a quantitative evaluation on a static dataset is limited in validity. While Offline RL methods could be explored with logged interactions, they are outside the scope of this work. Due to limited access to sustained live attack traffic, no statistically robust field evaluation has been completed to date. The following, therefore, documents the prototype's design and intended evaluation protocol rather than reporting conclusive empirical results. The planned protocol includes collecting deployment logs and performance metrics to compare RL-driven orchestration against non-RL baselines (e.g., threshold heuristics and static policies), followed by iterative refinement of the RL architecture.

\subsection{Experimental Setup}
An initial short-lived deployment was prepared to validate integration and data flows. However, the duration and volume of observed traffic were insufficient to derive statistically meaningful conclusions. A future, extended field study is outlined to assess detection quality, resource efficiency, and operational stability under representative load.

\subsubsection{Quantitative Results}
No quantitative results are reported at this time. Metrics and protocol for future evaluation are defined (precision/recall, F1, AUC; deployment efficiency; time-to-redirect; resource cost per engagement), with explicit comparison of RL versus non-RL baselines. Measurement is deferred until a sustained live trial is conducted.

\subsection{Discussion}
These observations highlight key areas for future work. The current RL agent and feature set are an initial operational design intended to exercise the orchestration loop. Demonstrating detection performance and selectivity requires an extended live trial with sufficient adversary engagement, which is part of the planned research agenda.

\section{Conclusion}
\label{sec:conclusion}

\noindent
This work addresses the critical limitations of static honeypots in countering modern, sophisticated cyber threats. A novel adaptive honeynet architecture is proposed that integrates reinforcement learning with dynamic container orchestration to create an intelligent and responsive cyber-deception system.

The primary contribution of this research is the design and implementation of a functional prototype that operationalizes the core architectural loop. While the RL agent (DQN with LSTM and dueling head) has been integrated end-to-end, empirical validation of detection quality and deployment selectivity remains future work pending a sustained live trial. An important operational lesson learned is that a deployment pipeline based on bespoke Bash scripts was too error-prone for repeatable operations and open-source scalability. As a result, deployment is being migrated to Ansible-based playbooks and roles to ensure idempotency, clearer state management, and maintainable contributions. The work provides a foundational blueprint and an initial, operational design for how RL can be leveraged for infrastructure-level orchestration.

A significant technical and methodological lesson emerged from the inability to fully realize the attacker forwarding mechanism as planned. As detailed in Section~\ref{sec:implementation_details}, the intended WireGuard-based tunneling for secure attacker traffic forwarding proved challenging to implement, requiring considerable effort without producing a functional result within the thesis timeframe. The subsequent availability of a DNAT-based alternative came too late to integrate into the operational pipeline, resulting in the absence of high-interaction (deep emulation) session data. These post-escalation logs are essential for evaluating the anomaly detection pipeline and implementing the envisioned quality-weighted reward function. Their absence limited the evaluation to first-flight decision-making and highlighted a broader dependency: in live cyber-defense experimentation, data availability is inseparably tied to the readiness and stability of critical infrastructure components. Delays or failures in implementing such core data-path elements can directly constrain the scope and validity of experimental outcomes.

The experience underlines the importance of prioritizing essential data collection mechanisms early in the development process, even before other architectural optimizations. For future iterations, attacker forwarding and post-escalation data capture will be treated as primary engineering milestones, ensuring that subsequent model evaluations are grounded in complete, representative datasets. While the prototype is subject to these limitations, it nevertheless serves as a crucial first step, proving the feasibility of the core concept and providing a robust platform for the extensive future work outlined in the next section. The transition to Ansible further reduces operational risk and opens the pathway for community-driven, reproducible deployments across diverse environments. A structured evaluation plan will collect and analyze data from deployments to compare the RL-driven orchestration loop against non-RL baselines, iteratively refine the agent architecture, and substantiate claims of superiority where supported by evidence. Ultimately, this research paves the way for a new class of proactive, autonomous defense systems capable of adapting to the ever-evolving threat landscape while maintaining an up-to-date catalog of bot families and versions.

\section{Future Work}
\label{sec:future_work}

\subsection{Security Enhancements}
\noindent
To further enhance the security of the adaptive honeynet, several key areas require attention. These enhancements aim to improve the system's resilience against sophisticated attacks and ensure robust protection mechanisms.

\subsubsection{Advanced Threat Detection}
Implementing advanced threat detection mechanisms will be critical for identifying and mitigating complex attack vectors. This will involve behavioral analysis to detect anomalies and potential threats by examining adversary activity patterns, signature-based detection using regularly updated threat intelligence to recognize known malicious activities, and heuristic analysis techniques to uncover previously unknown or zero-day threats.

\subsubsection{Enhanced Encryption}
Securing data in transit and at rest is vital to protecting sensitive information. Planned measures include enforcing Transport Layer Security (TLS) for all network communications to guarantee data integrity and confidentiality, as well as encrypting stored data with strong cryptographic algorithms to prevent unauthorized access.

Future efforts will be organized around three core pillars: enhancing the AI pipeline, building out advanced intelligence capabilities, and operationalizing the platform.

\subsection{Realizing the Architectural Vision}
A primary focus of future work will be the full implementation of the conceptual AI analytics pipeline, including the adaptive autoencoder for post-interaction log analysis and the anomaly-weighted reward function. This will transition the system from its current prototype stage to an intelligent data-gathering platform that prioritizes novel threats. Key research challenges include designing a robust online training process to counter concept drift~\cite{kirkpatrick2016overcoming-377} and exploring more advanced models like Transformers for log analysis. In parallel, a data-driven evaluation campaign will measure detection quality, deployment selectivity, time-to-redirect, and cost per engagement under controlled deployments, explicitly comparing the RL-driven orchestration to deterministic heuristics and static policies. Results will be used to refine the RL architecture (state, reward, and hyperparameters) and to determine where RL provides statistically significant advantages.

Furthermore, development will concentrate on building out the core intelligence capabilities outlined in the architecture, such as automated attack chain extraction, bot versioning, MITRE ATT\&CK integration, and incorporating Explainable AI (XAI) techniques~\cite{arrieta2020explainable-672} to ensure analyst trust and utility.

\subsection{Operationalization and Autonomous Capabilities}
The long-term vision is to develop a more autonomous system that reduces the need for constant human oversight. This includes research into \textbf{self-healing capabilities}, where the system can automatically recover from component failures or even direct attacks on the honeynet infrastructure. Furthermore, the RL agent's reward function and action space will be expanded to enable more complex \textbf{adaptive deception strategies}, where the honeypots themselves can modify their behavior in real-time to maximize intelligence gathering from high-value targets. Finally, the entire platform will be hardened and optimized for robust, scalable deployment in enterprise and national security environments. A staged rollout plan will introduce canary evaluations where RL policies are evaluated side-by-side with baseline policies to ensure safe, evidence-based progression.

\subsection{Scalability to a Global Honeynet}
    A critical next step is to evolve the prototype into a fully scalable, distributed system. This involves several concrete engineering efforts:
    \begin{itemize}
        \item \textbf{Multi-Sensor Deployment:} Replace ad-hoc bash scripts with Ansible playbooks and roles for reproducible, idempotent provisioning of sensors, Hive, and cluster nodes across diverse environments (on-prem and multiple clouds). Standardize inventories, variables, and secrets management to enable contributors to stand up testbeds reliably.
        \item \textbf{Hive and Cluster Federation:} For very large-scale deployments, the single Hive and Cluster model will become a bottleneck. Future research will explore a federated architecture with multiple, geographically distributed cluster nodes, each managed by a regional Hive. These regional Hives would, in turn, report high-level intelligence to a central "super-Hive" for global threat correlation.
        \item \textbf{Load Balancing and High Availability:} Implement load balancing for the Hive node's data ingestion endpoints and ensure high availability for the Elasticsearch cluster and RL agent to handle telemetry from a large fleet of sensors without data loss or downtime. Integrate Ansible with CI/CD to test HA failover playbooks and configuration drift remediation.
        \item \textbf{Optimized Data Telemetry:} Investigate and implement more efficient data serialization formats and communication protocols (e.g., gRPC with Protocol Buffers) for sensor-to-Hive communication to minimize bandwidth consumption, which is critical in a large-scale deployment.
        \item \textbf{Open-Source Workflow Enablement:} Provide an Ansible collection with modular roles (sensor, Hive, cluster, observability), plus Molecule tests, so external contributors can validate changes locally. Add GitHub Actions for linting (ansible-lint), idempotency checks, and basic integration smoke tests. Include datasets and scripts for reproducible evaluation runs to facilitate independent comparison of RL vs. non-RL baselines.
    \end{itemize}
\phantomsection
\section*{Acknowledgment}
The author would like to thank Heiko Folkerts and Detlef Nuß from the German Federal Office for Information Security (BSI) for their valuable support and collaboration on this project. The author is also grateful to Prof. Dr. Stephan Schneider and Prof. Dr. Christian Kraus from the University of Applied Sciences Kiel for their insightful guidance and academic supervision. 

The author further acknowledges the use of generative AI for assistance in refining the language, improving grammatical clarity, and supporting LaTeX formatting during the preparation of this manuscript. All research design, data analysis, interpretations, and conclusions remain entirely the responsibility of the author.

\section*{EU IA Act}
The authors acknowledge that the presented system incorporates AI-based decision-making in a cybersecurity context, bringing it within the scope of the European Union's AI Act (Regulation (EU) 2024/1689).~\cite{european-union2024ai-382} While the architecture does not meet the classification criteria for a"high-risk" AI system as defined in Annex III, its design intentionally aligns with the Act's general principles in Articles 10-15, including transparent feature extraction, auditable decision flows, and robust operation, to facilitate future compliance should the regulatory scope evolve.

\bibliographystyle{IEEEtran}
\bibliography{references}

\begin{IEEEbiography}[{\includegraphics[width=1in,height=1.25in,clip,keepaspectratio]{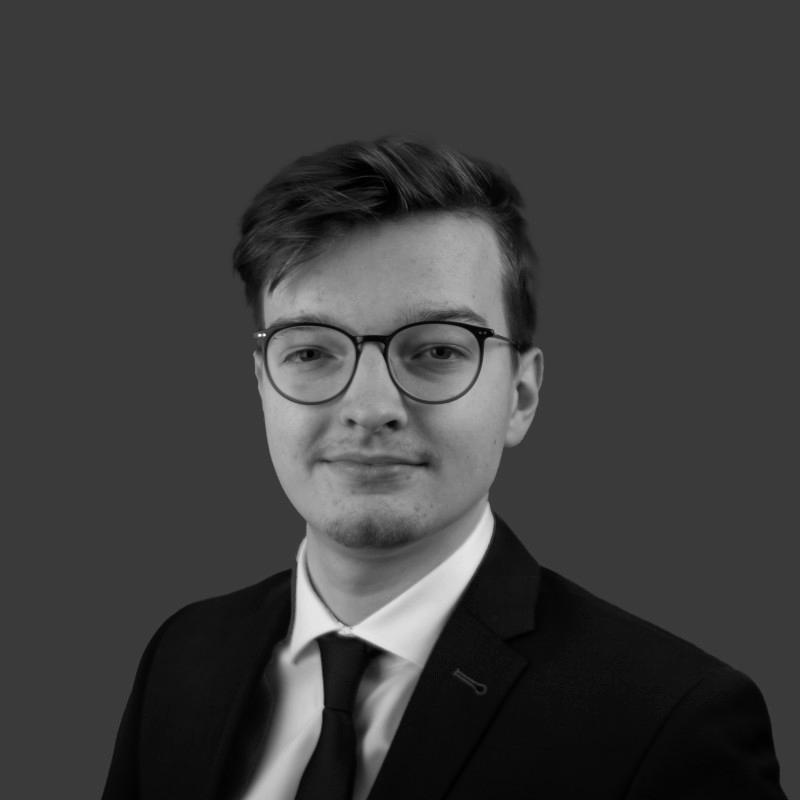}}]{Lukas Johannes Möller}
is currently a graduate student with the Georgia Institute of Technology, collaborating with the German Federal Office for Information Security (BSI) on adaptive honeynet and moving-target defense projects. His research interests include adaptive honeynets, reinforcement-learning-driven container orchestration, unsupervised anomaly detection, and vulnerability research.
\end{IEEEbiography}

\clearpage
\appendices
\onecolumn
\section{Detailed Feature Analysis of the MADCAT Sensor Data}
\label{app:appendix_featured_analysis}
\begin{longtable}{p{3cm} p{2cm} p{1.2cm} p{9cm}}
\toprule
\textbf{Feature} & \textbf{Data Type} & \textbf{Relevant} & \textbf{Justification} \\
\midrule
\endfirsthead
\toprule
\textbf{Feature} & \textbf{Data Type} & \textbf{Relevant} & \textbf{Justification} \\
\midrule
\endhead
\_index & object & No & Internal Elasticsearch metadata with no relation to session behavior. \\
\_id & object & No & Unique document identifier not related to network behavior. \\
\_score & object & No & Search relevance score unrelated to traffic features. \\
\_ignored & object & No & Field ingestion metadata unrelated to network behavior. \\
sort & object & No & Elasticsearch sort field not related to network features. \\
src\_ip & object & Yes & Direct source address is essential for identifying the origin of traffic in RL decisions. \\
dest\_ip\_geoip.geo.city\_name & object & No & City-level geolocation requires enrichment and is not reliable enough for direct RL scoring. \\
dest\_ip\_geoip.geo.region\_name & object & No & Region-level geolocation is enrichment data not directly usable in real-time RL decisions. \\
dest\_ip\_geoip.geo.location.lat & float64 & No & Latitude requires enrichment and is not used directly in RL decision-making. \\
dest\_ip\_geoip.geo.location.lon & float64 & No & Longitude requires enrichment and is not used directly in RL decision-making. \\
dest\_ip\_geoip.geo.continent\_code & object & No & Continent code is enrichment data and not necessary for RL scoring. \\
dest\_ip\_geoip.geo.country\_iso\_code & object & No & Country code requires enrichment and is not used in the RL scoring loop. \\
dest\_ip\_geoip.geo.postal\_code & object & No & Postal code is unreliable and irrelevant for RL scoring. \\
dest\_ip\_geoip.geo.country\_name & object & No & Human-readable country name is redundant and not needed for RL logic. \\
dest\_ip\_geoip.geo.timezone & object & No & Timezone is enrichment metadata and not a real-time RL feature. \\
TCP.res1 & float64 & No & Reserved header field with no behavioral value. \\
TCP.ack & object & Yes & ACK flag is directly usable for detecting connection state anomalies in RL. \\
TCP.syn & object & Yes & SYN flag is directly usable for detecting scans and session initiation patterns. \\
TCP.psh & object & No & PSH flag requires deep interpretation and is not a primary RL trigger. \\
TCP.urg & object & No & URG flag is rarely set and has no consistent RL scoring value. \\
TCP.rst & object & Yes & RST flag is directly usable for identifying scanning or forced connection termination. \\
TCP.src\_port & float64 & Yes & Source port is a fundamental feature for identifying service patterns. \\
TCP.tcp\_flags & object & Yes & Combined flags are directly usable for RL pattern matching. \\
TCP.ack\_seq & float64 & No & Acknowledgment sequence number requires session tracking beyond RL's real-time scope. \\
TCP.tcp\_options.tained & object & No & Tool-specific metadata unrelated to RL scoring. \\
TCP.tcp\_options.timestamp & object & No & TCP timestamp is enrichment-level fingerprinting, not used in RL. \\
TCP.tcp\_options.sack\_perm & object & No & SACK permission is fingerprinting data, not used in RL scoring. \\
TCP.tcp\_options.padding\_hex & object & No & Padding content is irrelevant for RL scoring. \\
TCP.tcp\_options.nop & object & No & No-Operation option has no RL scoring value. \\
TCP.tcp\_options.window & object & No & TCP window option requires enrichment and is not used in RL directly. \\
TCP.tcp\_options.mss & object & No & MSS is passive fingerprinting, not a real-time RL feature. \\
TCP.window & float64 & No & Window size is passive fingerprinting, not used in RL scoring. \\
TCP.cwr & object & No & Congestion flag is rare and irrelevant for RL. \\
TCP.urg\_ptr & object & No & Urgent pointer has no consistent RL scoring impact. \\
TCP.fin & object & Yes & FIN flag is directly usable for detecting unusual connection termination patterns. \\
TCP.dest\_port & float64 & Yes & Destination port is a fundamental RL decision variable. \\
TCP.hdr\_len & float64 & No & TCP header length is fingerprinting data, not used in RL decisions. \\
TCP.ecn & object & No & ECN flag is not a primary RL decision feature. \\
TCP.checksum & object & No & Checksum validity is handled at transport level, not used in RL scoring. \\
TCP.seq & float64 & No & Sequence number analysis is beyond the RL decision window. \\
src\_port & float64 & Yes & Source port at root level is usable for RL scoring if protocol-specific fields are missing. \\
FLOW.proxy\_ip & object & No & Proxy IP is enrichment data not available in real-time RL context. \\
FLOW.end & object & Yes & Flow end time is usable for session duration calculation in RL. \\
FLOW.start & object & Yes & Flow start time is directly usable for session duration in RL. \\
FLOW.bytes\_toclient & float64 & Yes & Bytes to client is directly usable for detecting exfiltration patterns in RL. \\
FLOW.backend\_ip & object & No & Backend IP requires enrichment and is not directly usable in RL scoring. \\
FLOW.proxy\_port & float64 & No & Proxy port is enrichment detail not used in RL scoring. \\
FLOW.duration & float64 & Yes & Flow duration is directly usable for RL attack behavior classification. \\
FLOW.bytes\_toserver & float64 & Yes & Bytes to server is directly usable for detecting suspicious uploads in RL. \\
FLOW.backend\_port & float64 & No & Backend port requires enrichment, not directly used in RL. \\
FLOW.reason & object & No & Flow termination reason is enrichment-level context, not a direct RL signal. \\
FLOW.min\_rtt & float64 & No & RTT measurement requires enrichment and is not part of RL scoring. \\
FLOW.state & object & Yes & Flow state is a direct RL decision feature. \\
src\_ip\_geoasn.as.number & float64 & Yes & ASN number of source IP is directly usable for RL reputation scoring. \\
src\_ip\_geoasn.as.organization.name & object & No & ASN organization name is enrichment data, not a direct RL variable. \\
log.offset & int64 & No & Log offset is ingestion metadata, irrelevant to RL. \\
log.file.path & object & No & Log file path is ingestion metadata, irrelevant to RL. \\
dest\_port & float64 & Yes & Destination port at root level is usable for RL scoring. \\
proto & object & Yes & Protocol type is a fundamental RL decision feature. \\
dest\_ip & object & Yes & Destination IP is directly usable for RL scoring. \\
ecs.version & object & No & ECS version is logging metadata. \\
observer.product & object & No & Observer product is logging metadata. \\
observer.version & object & No & Observer version is logging metadata. \\
observer.type & object & No & Observer type is logging metadata. \\
observer.hostname & object & Yes & Sensor hostname is usable for RL asset targeting logic. \\
xlog\_origin & object & No & Origin tag is metadata unrelated to RL scoring. \\
IP.version & float64 & Yes & IP version is directly usable for RL decisions. \\
IP.tot\_len & float64 & Yes & Total packet length is directly usable for RL anomaly scoring. \\
IP.tos & object & No & Type of Service is enrichment-level fingerprinting, not used in RL. \\
IP.dest\_addr & object & Yes & Destination address is directly usable for RL decisions. \\
IP.src\_addr & object & Yes & Source address is directly usable for RL decisions. \\
IP.id & object & No & IP ID is fingerprinting data, not used in RL scoring. \\
IP.hdr\_len & float64 & No & IP header length is fingerprinting data, not used in RL scoring. \\
IP.ttl & float64 & Yes & TTL is directly usable for RL OS/liveness detection. \\
IP.protocol & float64 & Yes & Protocol number is a fundamental RL decision feature. \\
IP.checksum & object & No & IP checksum is validation metadata, not a RL feature. \\
IP.flags & object & No & IP flags require enrichment to interpret and are not used directly in RL. \\
timestamp & object & Yes & Event timestamp is directly usable for RL decision timing. \\
timestamp\_processed & object & No & Processing timestamp is ingestion metadata. \\
backend\_id & object & No & Backend ID is infrastructure metadata. \\
SPLIT.split & object & No & Log splitting flag is ingestion metadata. \\
event.original & object & No & Raw log text is not parsed for direct RL use. \\
src\_ip\_geoip.geo.country\_name & object & No & Source country name is enrichment data not used directly in RL. \\
src\_ip\_geoip.geo.location.lat & float64 & No & Source latitude is enrichment data, not a RL feature. \\
src\_ip\_geoip.geo.location.lon & float64 & No & Source longitude is enrichment data, not a RL feature. \\
src\_ip\_geoip.geo.continent\_code & object & No & Continent code is enrichment data, not used in RL scoring. \\
src\_ip\_geoip.geo.country\_iso\_code & object & No & Country code is enrichment data, not used in RL scoring. \\
src\_ip\_geoip.geo.timezone & object & No & Timezone is enrichment data, not used in RL scoring. \\
@version & object & No & Logstash metadata. \\
origin & object & No & Log origin metadata. \\
host.containerized & bool & Yes & Containerization state is directly usable for RL targeting logic. \\
host.mac & object & No & MAC address is not visible for most remote sessions. \\
host.name & object & Yes & Hostname is directly usable for RL targeting logic. \\
host.architecture & object & No & CPU architecture requires enrichment, not used in RL scoring. \\
host.ip & object & Yes & Host IP is directly usable for RL targeting logic. \\
host.id & object & No & Host ID is internal metadata. \\
host.os.version & object & No & OS version is enrichment data, not used in RL scoring. \\
host.os.kernel & object & No & Kernel version is enrichment data, not used in RL scoring. \\
host.os.family & object & No & OS family is enrichment data, not used in RL scoring. \\
host.os.name & object & No & OS name is enrichment data, not used in RL scoring. \\
host.os.codename & object & No & OS codename is enrichment data, not used in RL scoring. \\
host.os.platform & object & No & OS platform is enrichment data, not used in RL scoring. \\
host.os.hostname & object & Yes & OS hostname is directly usable for RL targeting logic. \\
input.type & object & No & Input type is ingestion metadata. \\
ct\_status & object & Yes & Threat intelligence status is directly usable for RL blocking logic. \\
unixtime & float64 & Yes & Unix timestamp is directly usable for RL timing logic. \\
@timestamp & object & Yes & Event timestamp is directly usable for RL decisions. \\
event\_type & object & Yes & Event type is directly usable for RL decision-making. \\
agent.version & object & No & Agent version is metadata. \\
agent.name & object & No & Agent name is metadata. \\
agent.id & object & No & Agent ID is metadata. \\
agent.type & object & No & Agent type is metadata. \\
agent.hostname & object & No & Agent hostname is redundant metadata. \\
agent.ephemeral\_id & object & No & Agent ephemeral ID is metadata. \\
tags & object & No & Tags require enrichment parsing, not direct RL use. \\
dest\_ip\_geoasn.as.number & float64 & Yes & Destination ASN number is directly usable for RL reputation scoring. \\
dest\_ip\_geoasn.as.organization.name & object & No & Destination ASN org name is enrichment data. \\
FLOW.payload\_hd & object & No & Hex payload dump is not parsed in real-time RL. \\
FLOW.payload\_str & object & No & String payload is not parsed in real-time RL. \\
FLOW.payload\_sha1 & object & Yes & Payload hash is directly usable for RL matching to known threats. \\
host.os.type & object & No & OS type is enrichment data, not used in RL scoring. \\
src\_ip\_geoip.mmdb.dma\_code & float64 & No & DMA code is irrelevant for RL scoring. \\
src\_ip\_geoip.geo.city\_name & object & No & City name is enrichment data, not used in RL scoring. \\
src\_ip\_geoip.geo.postal\_code & object & No & Postal code is unreliable and irrelevant for RL scoring. \\
src\_ip\_geoip.geo.region\_name & object & No & Region name is enrichment data, not used in RL scoring. \\
TCP.payload\_hd & object & No & Hex TCP payload is not parsed for RL. \\
TCP.payload\_sha1 & object & Yes & TCP payload hash is directly usable for RL threat matching. \\
TCP.payload\_str & object & No & String TCP payload is not parsed for RL. \\
TCP.tcp\_options.eol & object & No & End of Option List is irrelevant to RL scoring. \\
icmp\_code & float64 & Yes & ICMP code is directly usable for RL anomaly detection. \\
ICMP.type\_str & object & Yes & ICMP type string is directly usable for RL scoring. \\
ICMP.id & object & No & ICMP ID is not a primary RL feature. \\
ICMP.checksum & object & No & ICMP checksum is irrelevant to RL scoring. \\
ICMP.code & float64 & Yes & ICMP code is directly usable for RL scoring. \\
ICMP.tainted & object & No & Tool-specific metadata irrelevant to RL. \\
ICMP.type & float64 & Yes & ICMP type is directly usable for RL anomaly scoring. \\
ICMP.seq & object & No & ICMP sequence number is not a primary RL feature. \\
icmp\_type & float64 & Yes & ICMP type number is directly usable for RL scoring. \\
http.version & object & No & HTTP version is enrichment-level metadata. \\
http.request.method & object & Yes & HTTP method is directly usable for RL web attack detection. \\
user\_agent.original & object & No & Full user agent string requires parsing, not direct RL use. \\
user\_agent.os.full & object & No & OS info from UA is enrichment data, not RL-direct. \\
user\_agent.os.name & object & No & OS name from UA is enrichment data. \\
user\_agent.name & object & No & UA name is enrichment data. \\
user\_agent.device.name & object & No & UA device type is enrichment data. \\
UDP.len & float64 & Yes & UDP length is directly usable for RL anomaly scoring. \\
UDP.src\_port & float64 & Yes & UDP source port is directly usable for RL scoring. \\
UDP.checksum & float64 & No & UDP checksum is irrelevant to RL scoring. \\
UDP.dest\_port & float64 & Yes & UDP destination port is directly usable for RL scoring. \\
url.original & object & Yes & Full URL is directly usable for RL web attack detection. \\
user\_agent.version & object & No & UA version is enrichment data. \\
RAW.ether\_type & object & No & EtherType is L2 metadata irrelevant to RL scoring. \\
RAW.pcap\_filter & object & No & Capture filter is collection metadata. \\
tainted & object & No & Generic taint flag is irrelevant to RL. \\
SPLIT.part & float64 & No & Split part number is ingestion metadata. \\
SPLIT.total & float64 & No & Split total is ingestion metadata. \\
SPLIT.tag & object & No & Split tag is ingestion metadata. \\
ICMP.IP.version & float64 & No & Encapsulated IP version is enrichment detail. \\
ICMP.IP.tos & object & No & Encapsulated TOS is enrichment detail. \\
ICMP.IP.tot\_len & float64 & No & Encapsulated total length is enrichment detail. \\
ICMP.IP.dest\_addr & object & No & Encapsulated dest address is enrichment detail. \\
ICMP.IP.src\_addr & object & No & Encapsulated source address is enrichment detail. \\
ICMP.IP.id & object & No & Encapsulated IP ID is enrichment detail. \\
ICMP.IP.hdr\_len & float64 & No & Encapsulated header length is enrichment detail. \\
ICMP.IP.ttl & float64 & No & Encapsulated TTL is enrichment detail. \\
ICMP.IP.protocol & float64 & No & Encapsulated protocol is enrichment detail. \\
ICMP.IP.flags & object & No & Encapsulated flags are enrichment detail. \\
ICMP.IP.checksum & object & No & Encapsulated checksum is irrelevant. \\
ICMP.TCP.res1 & float64 & No & Reserved encapsulated TCP field is irrelevant. \\
ICMP.TCP.ack & object & No & Encapsulated ACK is enrichment detail. \\
ICMP.TCP.psh & object & No & Encapsulated PSH is enrichment detail. \\
ICMP.TCP.syn & object & No & Encapsulated SYN is enrichment detail. \\
ICMP.TCP.urg & object & No & Encapsulated URG is enrichment detail. \\
ICMP.TCP.src\_port & float64 & No & Encapsulated source port is enrichment detail. \\
ICMP.TCP.tcp\_flags & object & No & Encapsulated flags are enrichment detail. \\
ICMP.TCP.rst & object & No & Encapsulated RST is enrichment detail. \\
ICMP.TCP.ack\_seq & float64 & No & Encapsulated ACK seq is enrichment detail. \\
ICMP.TCP.window & float64 & No & Encapsulated window is enrichment detail. \\
ICMP.TCP.cwr & object & No & Encapsulated CWR is enrichment detail. \\
ICMP.TCP.urg\_ptr & object & No & Encapsulated urgent pointer is enrichment detail. \\
ICMP.TCP.fin & object & No & Encapsulated FIN is enrichment detail. \\
ICMP.TCP.dest\_port & float64 & No & Encapsulated destination port is enrichment detail. \\
ICMP.TCP.hdr\_len & float64 & No & Encapsulated header length is enrichment detail. \\
ICMP.TCP.ecn & object & No & Encapsulated ECN is enrichment detail. \\
ICMP.TCP.checksum & object & No & Encapsulated checksum is irrelevant. \\
ICMP.TCP.seq & float64 & No & Encapsulated sequence number is enrichment detail. \\
ICMP.unused & object & No & ICMP unused field is irrelevant. \\
ICMP.code\_str & object & Yes & Human-readable ICMP code is directly usable in RL scoring. \\
ICMP.TCP.tcp\_options.tained & object & No & Tool-specific encapsulated metadata is irrelevant. \\
ICMP.TCP.tcp\_options.mss & object & No & Encapsulated MSS is enrichment detail. \\
ICMP.TCP.tcp\_options.padding\_hex & object & No & Encapsulated padding is irrelevant. \\
http.request.referrer & object & No & HTTP referrer is enrichment-level data, not a direct RL feature. \\
ICMP.TCP.tcp\_options.nop & object & No & Encapsulated NOP is irrelevant. \\
ICMP.TCP.tcp\_options.timestamp & object & No & Encapsulated TCP timestamp is enrichment detail. \\
\bottomrule
\end{longtable}
\EOD
\end{document}